\documentclass[fleqn,usenatbib]{mnras}

\usepackage{color}
\usepackage[T1]{fontenc}
\usepackage{ae,aecompl}
\usepackage{xcolor}
\usepackage{ulem}

\usepackage{graphicx}	
\usepackage{amsmath}	
\usepackage{cases}
\usepackage{adjustbox}

\usepackage{graphicx}	
\usepackage{amssymb}	
\usepackage{newtxtext,newtxmath}
\usepackage{cases}
\usepackage{array,multirow}
\usepackage{upgreek}
\usepackage{textcomp}
\usepackage{rotating}
\usepackage{fontawesome}

\newcommand{\Msol}{\,{\rm M}_{\odot}}
\newcommand{\Rsol}{\,{\rm R}_{\odot}}
\newcommand{\gram}{\;\mathrm{g}}
\newcommand{\cm}{\;\mathrm{cm}}

\newcommand{\pc}{\;\mathrm{pc}}
\newcommand{\yr}{\;\mathrm{yr}}
\newcommand{\mstar}{\;M_{\star}}
\newcommand{\rstar}{\;R_{\star}}
\newcommand{\Mbh}{M_{\bullet}}

\title[TDEs in AGNs]{In-plane Tidal Disruption of Stars in Disks of Active Galactic Nuclei }

\author[T. Ryu et al.]{
Taeho Ryu$^{1,2}$\thanks{E-mail: tryu@mpa-garching.mpg.de}, Barry McKernan$^{3,4,5,6}$, K.E. Saavik Ford$^{3,4,5,6}$, Matteo Cantiello$^{3}$,
\newauthor Matthew Graham$^{7}$, Daniel Stern $^{8}$, Nathan W.C. Leigh$^{9,4}$\\
$^{1}$Max Planck Institute for Astrophysics, Karl-Schwarzschild-Str.~1, 85748 Garching, Germany\\%
$^{2}$Physics and Astronomy Department, Johns Hopkins University, Baltimore, MD 21218, USA\\%
$^{3}$Center for Computational Astrophysics, Flatiron Institute, New York, NY 10010, USA\\%
$^{4}$Department of Astrophysics, American Museum of Natural History, New York, NY 10024, USA\\%
$^{5}$Graduate Center, City University of New York, 365 5th Avenue, New York, NY 10016, USA\\%
$^{6}$Department of Science, BMCC, City University of New York, New York, NY 10007, USA\\%
$^{7}$California Institute of Technology, 1200 E. California Blvd, Pasadena, CA 91125, USA\\%
$^{8}$Jet Propulsion Laboratory, California Institute of Technology, 4800 Oak Grove Drive, Pasadena, CA 91109, USA\\%
$^{9}$Departamento de Astronom\'a, Facultad de Ciencias F\'sicas y Matem\'aticas, Universidad de Concepci\'on, Concepci\'on, Chile \\%
}

\date{Accepted XXX. Received YYY; in original form ZZZ}

\pubyear{2022}

\begin{document}
\label{firstpage}
\pagerange{\pageref{firstpage}--\pageref{lastpage}}
\maketitle

\begin{abstract}
Stars embedded in active galactic nucleus (AGN) disks or captured by them may scatter onto the supermassive black hole (SMBH), leading to a tidal disruption event (TDE). Using the moving-mesh hydrodynamics simulations with {\small AREPO}, we investigate the dependence of debris properties in in-plane TDEs in AGN disks on the disk density and the orientation of stellar orbits relative to the disk gas (pro- and retro-grade). Key findings are: 1) Debris experiences continuous perturbations from the disk gas, which can result in significant and continuous changes in debris energy and angular momentum compared to `naked' TDEs. 2) Above a critical density of a disk around a SMBH with mass $M_{\bullet}$ ($\rho_{\rm crit} \sim 10^{-8}{\rm g~cm^{-3}}(M_{\bullet}/10^{6}{\rm M}_{\odot})^{-2.5}$) for retrograde stars, both bound and unbound debris is fully mixed into the disk. The density threshold for no bound debris return, inhibiting the accretion component of TDEs, is $\rho_{\rm crit,bound} \sim 10^{-9}{\rm g~cm^{-3}}(M_{\bullet}/10^{6}\Msol)^{-2.5}$. 3) Observationally, AGN-TDEs transition from resembling naked TDEs in the limit of $\rho_{\rm disk}\lesssim 10^{-2}\rho_{\rm crit,bound}$ to fully muffled TDEs with associated inner disk state changes at $\rho_{\rm disk}\gtrsim\rho_{\rm crit,bound}$, with a superposition of AGN+TDE in between. Stellar or remnant passages themselves can significantly perturb the inner disk. This can lead to an immediate X-ray signature and optically detectable inner disk state changes, potentially contributing to the changing-look AGN phenomenon. 4) Debris mixing can enriches the average disk metallicity over time if the star’s metallicity exceeds that of the disk gas. We point out signatures of AGN-TDEs may be found in large AGN surveys.
\end{abstract}



\section{Introduction}\label{sec:intro}

Active galactic nuclei (AGN) are powered by the accretion of gas disks onto supermassive black holes (SMBH). The accreting SMBH is often also orbited by a nuclear star cluster \citep{Neumayer+2020}, which must interact with the gas disk. Depending on the radial size of the gas disk, gas density ($\rho_{\rm disk}$), and how long the disk lasts, some fraction of the nuclear star cluster orbiting the SMBH will be captured by the AGN disk \citep[e.g.][]{Arty93,Fabj20,Nasim23,Generozov23,Yihan23}. Star formation within the AGN disk \citep[e.g.][]{GoodmanTan04,Levin07} can also add to the embedded stellar population. 
Thus, we expect a dynamic population of embedded objects (stars and stellar remnants) to live within AGN disks. The initial population of objects within the AGN disk soon after it forms should consist of both prograde and retrograde orbiters, leading to the possibility of dynamically complex and high-speed encounters and scatterings \citep[e.g.][]{Leigh18,Yihan21}. Captured orbiters may include stars on retrograde orbits and high eccentricities at small semi-major axes \citep{Yihan23}. In-plane tidal disruption events in AGN (or simply AGN-TDEs hereafter) can result from either in-plane scatterings of stars onto the SMBH, or eccentricity pumping of stars on retrograde orbits  \citep[e.g.][]{Secunda+2021,McKernan+2022}. Note that this is \textit{a new source} of TDEs in addition to standard loss-cone filling scattering yielding TDEs at roughly the same rate ($\sim 10^{-4} {\rm yr}^{-1}$) as in any other (quiescent) galactic nucleus \citep[e.g.,][]{Stone+2020}. The loss-cone TDEs \citep[e.g.,][]{Hills1988,Rees1988} will very likely intersect the AGN disk at an angle and yield a TDE that looks different from a TDE in a vacuum \citep{Chan+2019,Chan+2020,Chan21}. 

In-plane AGN TDEs should look more different still. In \citet{McKernan+2022} we speculated that AGN-TDEs could look quite different from `naked' or gas-free TDEs, with observable differences between TDEs that are retrograde or prograde compared to the flow of disk gas. Here we investigate the hydrodynamics of  prograde and retrograde AGN-TDEs using a simple disk model, with a view to qualitatively describing key features of AGN-TDEs and potential observables. Throughout we highlight the point that in-plane AGN TDEs test the dynamics of the disk as well as its embedded population. While stars on prograde orbits embedded in AGN disks may experience runaway mass growth \citep{Cantiello+2021,Jermyn+22}, we only consider TDEs of normal main-sequence stars.

The paper is organized as follows: we provide descriptions of our numerical methods in detail in \S\ref{sec:method}. We present results of our simulations in \S\ref{sec:results} and discuss astrophysical implications in \S\ref{sec:discussion} and caveats in \S\ref{sec:caveats}. Finally, we conclude with a summary in \S\ref{sec:conclusion}.

\section{Simulation details}\label{sec:method}

\subsection{Numerical methods}

We perform 3D hydrodynamic simulations of a tidal disruption event of a main-sequence (MS) star on an in-plane parabolic orbit\footnote{For naked TDEs, it is a good approximation that a star that is tidally disrupted initially had approached on a parabolic orbit. Even in AGNs, if a star approaches the SMBH from a large distance near the influence radius of the SMBH, the stellar orbit can be approximated as a parabolic orbit. The disk gas would exert a drag force on the star, affecting the orbit. But, in this work, we simply assume a parabolic orbit for simplicity.} around an AGN disk surrounding a SMBH, using the massively parallel gravity and magnetohydrodynamic moving-mesh code {\small AREPO} \citep{Arepo,ArepoHydro,Arepo2}. It employs a second-order finite-volume scheme to solve the hydrodynamic equations on a moving Voronoi mesh, and a tree-particle-mesh method for gravity calculations. By adopting this innovative approach to grid construction and solving hydrodynamics equations, the code inherits advantages of both commonly used hydrodynamics schemes, Eulerian grid-based methods and Lagrangian smoothed particle methods. The advantages include improved shock capturing without introducing an artificial viscosity, and adaptive adjustment of spatial resolution. 

We use the ideal gas equation $p=(\gamma - 1)u$ with $\gamma = 5/3$, where $p$ is the pressure and $u$ is the gas internal energy density.

\subsection{Creation of a star in an AGN disk}\label{sub:modeling}

To make proper initial conditions for the simulations, we follow several steps, 1) creating a disk (\S\ref{subsubsec:disk}), 2) creating a 3D MS star (\S\ref{subsubsec:star}), and 3) placing the star in the disk with a different mid-plane density on a parabolic orbit around a SMBH (\S\ref{subsubsec:initial}).
.

\subsubsection{AGN disk around a supermassive black hole}\label{subsubsec:disk}

We model the central SMBH using a non-rotating sink particle, which interacts solely through gravitational forces with the gas. We allow the particle to grow in mass via accretion following the same procedure described in \citet{Ryu+2023}. However, it's worth noting that the total mass accreted remains significantly smaller than the mass of the SMBH throughout the simulation. Consequently, the change in the gravitational potential due to the mass growth of the SMBH would not significantly impact the results presented in this paper. 

For the sake of completeness, we will briefly summarize the adopted accretion prescription. At every time step, the accretion rate is estimated as an inward radial mass flux towards the BH averaged over cells with weights within $10r_{\rm g}$ (denoted by ``accretion radius''), and multiplied by the integration area. Here, $r_{\rm g}=G\Mbh/c^{2}$ represents the gravitational radius, which is approximately $2\Rsol$ for a SMBH mass of $\Mbh=10^{6}\Msol$. The weights are given using an inverse-distance weighted spline Kernel \citep{MonaghanLattanzio1985} (Equation 4 in \citealt{GADGET2}). If there are only a few cells within the accretion radius, the accretion rate estimate may be affected by Poisson noise. To ensure a sufficient number of cells in proximity to the black hole, we dynamically adjust cell refinement and derefinement within a region slightly larger than the accretion radius, aiming to maintain more than approximately 100 cells within this radius. Specifically, the code refines cells with a density greater than $10^{-14}$ g/cm$^{3}$ and a mass exceeding $6\times10^{22}$ g, provided that the ratio of the cell size to the distance from the black hole is greater than $\Delta r/r = 0.03$. Conversely, the code derefines cells if their mass falls below $1.5\times10^{22}$ g or if $\Delta r/r < 0.01$.

\begin{figure}
    \centering
    \includegraphics[width=8.4cm]{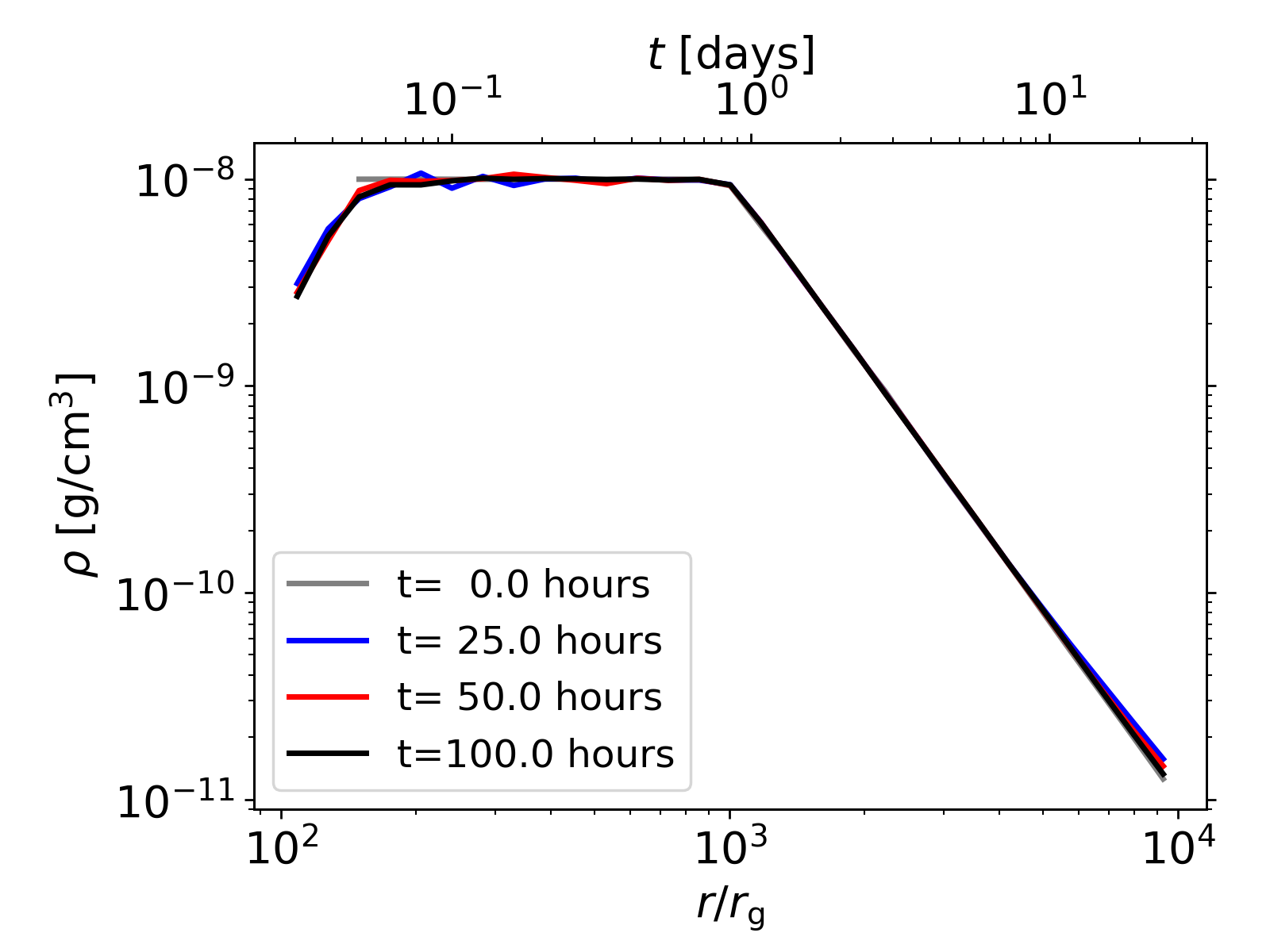}
\caption{Average mid-plane density and enclosed mass of our fiducial AGN disk model at four different times: $t=0$ hours (grey dotted), $25$ hours (blue solid), $50$ hours (red solid), $100$ hours (black solid). The upper $x-$axis indicates the time which it takes for a star on a parabolic orbit to reach the given distance.  }
	\label{fig:relaxeddisk1}
\end{figure}

\begin{figure}
    \centering
    \includegraphics[width=8.4cm]{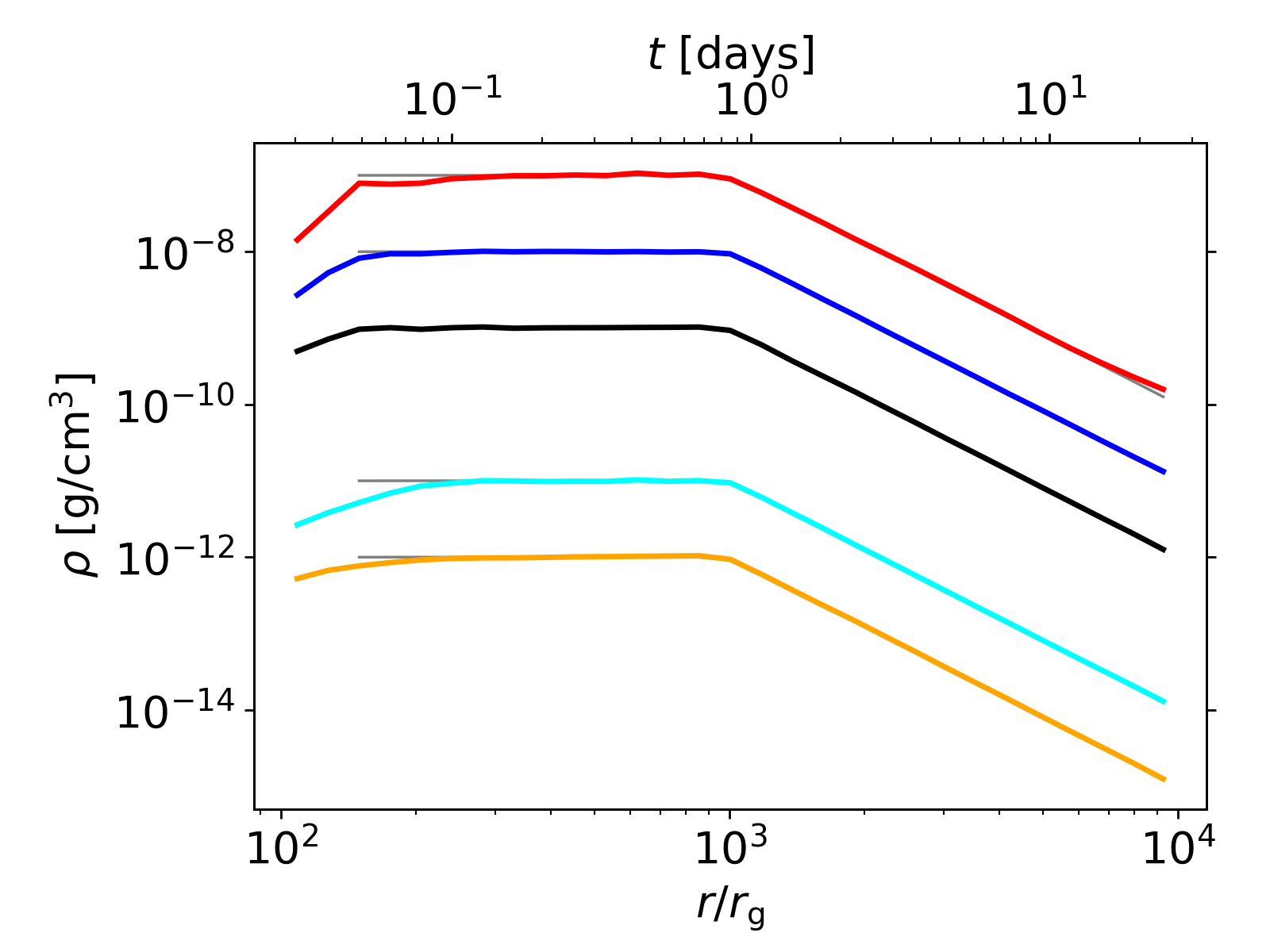}\\
    \includegraphics[width=8.4cm]{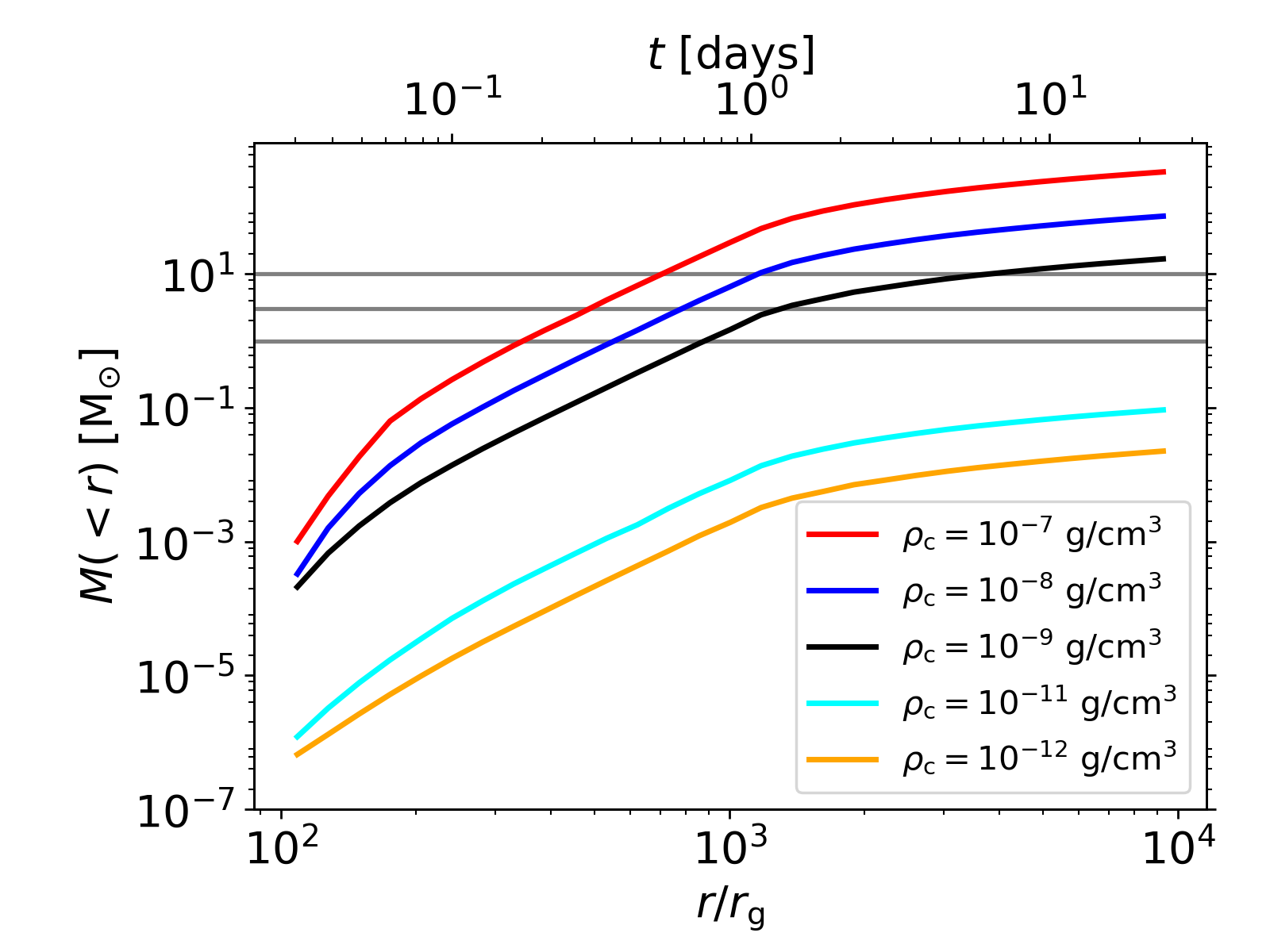}
    \caption{Mid-plane density (\textit{top}) and enclosed mass (\textit{bottom}) of fully relaxed disks with a different initial mid-plane density. The grey lines in the \textit{top} panel showing the initial density profile are just sitting on top of the profile for the relaxed disks. On the other hand, the grey horizontal lines in the \textit{bottom} panel indicate the masses of the stars considered. Like Figure~\ref{fig:relaxeddisk1}, the upper $x-$axis shows how long it takes for a star on a parabolic orbit to travel the given distance.   }
	\label{fig:relaxeddisk2}
\end{figure}

\begin{table*}
\begin{tabular}{ c c c c c c c c } 
\hline
$M_{\bullet}~[\Msol]$ & $\dot{M}/\dot{M}_{\rm Edd}$ & $\rho_{\rm c}$ [$\gram/\cm^{-3}$] & $T_{\rm c}$[K] & $h/r (r=R_{\rm inner})$ \\
\hline
\hline
$10^{6}$  & 0.5  & $10^{-7}$ & $8\times10^{7}$  & 0.04\\
$10^{6}$  & 0.5  & $10^{-8}$  & $4\times10^{8}$ & 0.08\\
$10^{6}$ & 0.5   & $10^{-9}$  & $2\times10^{9}$  & 0.16\\
$10^{6}$ & 0.001   & $10^{-11}$  & $6\times10^{8}$  & 0.10\\
$10^{6}$ & 0.001   & $10^{-12}$  & $3\times10^{9}$  & 0.21\\
\hline
\end{tabular}
\caption{ Disk parameters: (\textit{first}) black hole mass $\Mbh$, (\textit{second}) assumed mass accretion rate $\dot{M}$ relative to the Eddington mass accretion rate $M_{\rm Edd}$, (\textit{third}) maximum density in the mid-plane $\rho_{\rm c}$ (see Equation~\ref{eq:rho}), (\textit{fourth}) maximum temperature in the mid-plane $T_{\rm c}$ (see Equation~\ref{eq:T}), and (\textit{fifth}) aspect ratio at $r=R_{\rm inner}$. }\label{tab:initialparameter}
\end{table*}

\begin{table*}
\begin{tabular}{ c c c c c c c c } 
\hline
Model number & $M_{\bullet}~[\Msol]$ & $\mstar~[\Msol]$ & $\rho_{\rm c}$ [$\gram/\cm^{-3}$] & Pro. or retro. & $r_{\rm p}/r_{\rm t}$ & $r_{\rm p} [r_{\rm g}]$ & $t_{\rm p}$ [hours] \\
\hline
\hline
1 & $10^{6}$ & 1 & $10^{-7}$ & Pro & 0.3 & 13 & 0.07 \\
2 & $10^{6}$ & 1 & $10^{-7}$ & Retro & 0.3 & 13 &  0.07 \\
3 & $10^{6}$ & 1 & $10^{-8}$ & Pro & 0.3& 13 &  0.07 \\
4 & $10^{6}$ & 1 & $10^{-8}$ & Retro & 0.3 & 13 &  0.07 \\
5 & $10^{6}$ & 1 & $10^{-9}$ & Pro & 0.3 & 13 &  0.07 \\
6 & $10^{6}$ & 1 & $10^{-9}$ & Retro & 0.3 & 13 &  0.07 \\
7 & $10^{6}$ & 1 & $10^{-11}$ & Pro & 0.3 & 13 &  0.07 \\
8 & $10^{6}$ & 1 & $10^{-11}$ & Retro & 0.3 & 13 &  0.07 \\
9 & $10^{6}$ & 1 & $10^{-12}$ & Pro & 0.3 & 13 &  0.07 \\
10 & $10^{6}$ & 1 & $10^{-12}$ & Retro & 0.3 & 13 &  0.07 \\

\hline
11 & $10^{6}$ & 3 & $10^{-8}$ & Pro & 0.3 & 24 &  0.15 \\
12 & $10^{6}$ & 3 & $10^{-8}$ & Retro & 0.3 & 24 &  0.15 \\
\hline
13 & $10^{6}$ & 10 & $10^{-8}$ & Pro & 0.3& 30 &  0.22 \\
14 & $10^{6}$ & 10 & $10^{-8}$ & Retro & 0.3 & 30 &  0.22 \\
\hline
\end{tabular}
\caption{ Initial parameters: (left to right) model number, black hole mass $\Mbh [\Msol]$, stellar mass $\mstar[\Msol]$, maximum mid-plane density $\rho_{\rm c}$, relative orientation (prograde vs retrograde), pericenter distance $r_{\rm p}$ measured in units of the tidal radius $r_{\rm t}$, 
 $r_{\rm p}$ measured in units of the gravitational radius $r_{\rm g}$, dynamical time $t_{\rm p}$ at pericenter. }\label{tab:models}
\end{table*}

\begin{figure}
	\centering
	\includegraphics[width=8.4cm]{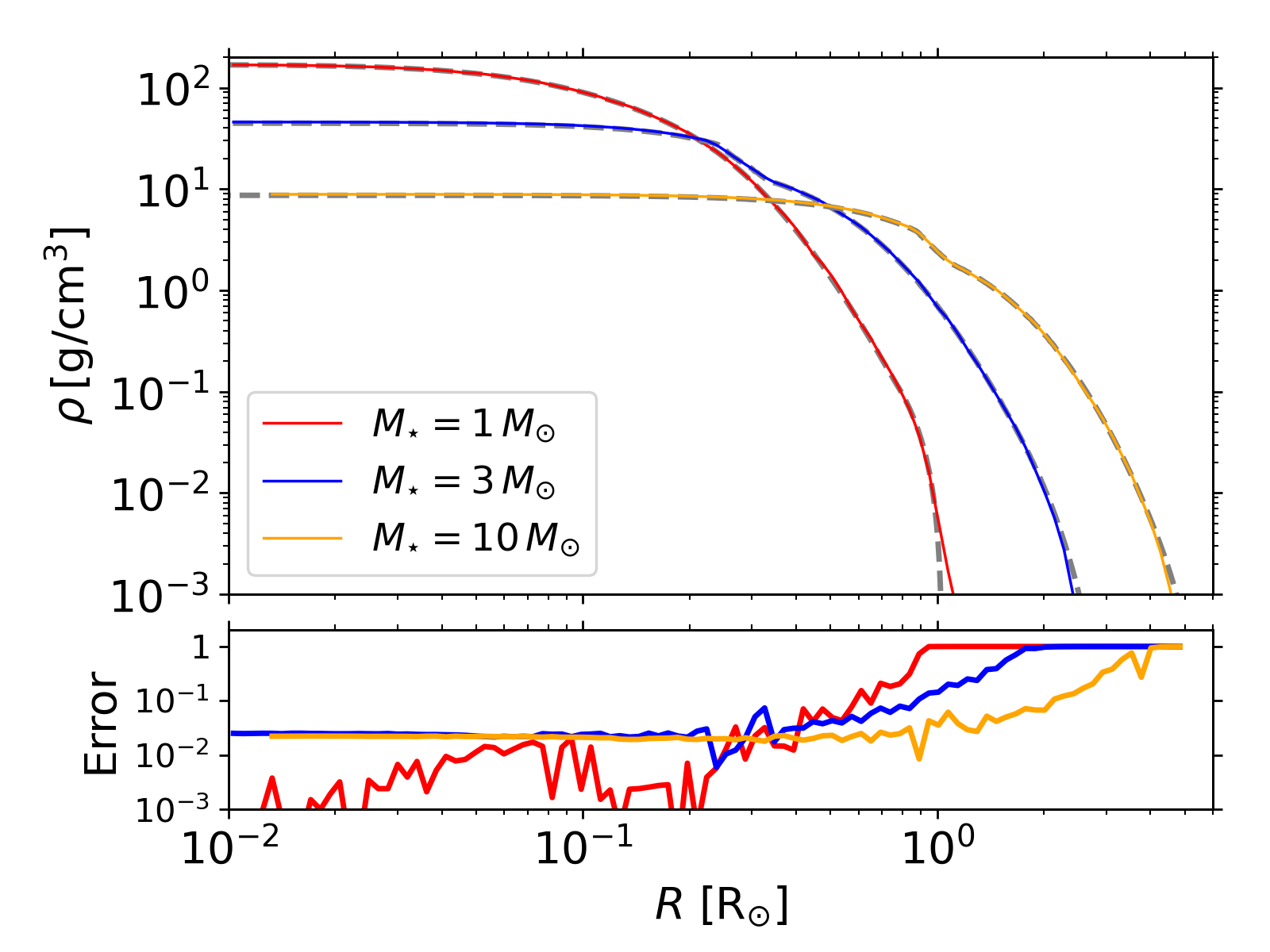}
\caption{Radial density (\textit{top}) of fully relaxed 3D stars with mass $M_{\star}=1\Msol$ (red solid), $3\Msol$ (blue solid), and $10\Msol$ (orange solid), over-plotted with lines for the MESA models (grey dashed lines) and the relative errors (\textit{bottom}) between the two density profiles.  }
	\label{fig:relaxedstar}
\end{figure}

The thermodynamic profiles of an AGN disk surrounding the SMBH are described by the solution for a gas-pressure dominated disk in \citet{Nelson+2013}. The mid-plane density and temperature follow a power-law in $r$,
\begin{align}\label{eq:rho}
    \rho_{\rm mid}(r) = \rho_{\rm c} \left(\frac{r}{r_{\rm cusp}}\right)^{p},\\\label{eq:T}
    T_{\rm mid}(r) = T_{\rm c} \left(\frac{r}{r_{\rm cusp}}\right)^{q},
\end{align}
where $\rho_{\rm c}$ and $T_{\rm c}$ are the mid-plane density and temperature near the inner edge of the disk, respectively, and $r_{\rm cusp} = 10^{3}r_{\rm g}$. 
In this work, we consider a disk surrounding a $10^{6}\Msol$ SMBH with the mid-plane density $\rho_{\rm c}=10^{-7}-10^{-12}$ g~cm$^{-3}$ at the inner disk edge $R_{\rm inner} = 100 r_{\rm g}$ (see Table~\ref{tab:models}). To match the disk solution by \citet{SirkoGoodman2003}, we adopt the values of $p$, 
\begin{align}
    p &= \begin{cases}
      & 0   \hspace{0.18in} {\rm for} ~r< r_{\rm cusp}\\
      & -3   \hspace{0.1in} {\rm for} ~r> r_{\rm cusp},
    \end{cases}
\end{align}
and $q =-3/4$. 

The vertical structure, i.e., the density and angular frequency $\Omega$, of the disk is described by the following equations,
\begin{align}
    \rho_{\rm disk}(r, z) &= \rho_{\rm mid}(r) \exp\left(\left[\frac{h}{r}\right]^{-2}\left[\frac{1}{\sqrt{1 + (z/r)^{2}}} - 1\right]\right),\\
    \Omega_{\rm disk}(r,z) &= \Omega_{\rm K}\left[(p+q) \left(\frac{h}{r}\right)^{2} + ( 1+ q) - \frac{q}{\sqrt{1 + (z/r)^{2}}}\right]^{1/2},
\end{align}
where $\Omega_{\rm K} = \sqrt{G\Mbh/r^{3}}$ is the Keplerian angular frequency and $h/r$ the aspect ratio. Note that the temperature has no dependence on the vertical distance orthogonal to the mid-plane $z$, meaning the temperature (so the sound speed) is constant along each vertical column at given $r$.

The disk is fully described if  $\rho_{\rm c}$, $T_{\rm c}$ and $h/r$ at $R_{\rm inner}$ are determined. For a given $\rho_{\rm c}$ we estimate the two other disk parameters using the following equations from \citet{SirkoGoodman2003},
\begin{align}\label{eq:1}
   c_{\rm s}^{2} \Sigma &= \frac{\dot{M}'\Omega}{3\pi \alpha},\\\label{eq:2}
   \Sigma &= 2 \rho_{\rm c} h,\\\label{eq:3}
   h &= \frac{c_{\rm s}}{\Omega},
\end{align}
where $\dot{M}' = \dot{M} (1 - \sqrt{\frac{5r_{\rm g}}{R_{\rm inner}}})$ and $\dot{M}$ is the accretion rate. Assuming $\dot{M}=0.5\dot{M}_{\rm Edd}$, $\dot{M}'\simeq 0.4 \dot{M}_{\rm Edd}$ at $R_{\rm inner}=100r_{\rm g}$. Here, $\dot{M}_{\rm Edd}=10L_{\rm Edd}/c^{2}$ where $L_{\rm Edd}$ is the Eddington luminosity and an radiation efficiency of $\eta=0.1$ is assumed. And $\alpha = 0.01$ is the viscosity parameter. Combining Equations \ref{eq:1}, \ref{eq:2}, and \ref{eq:3}, we find an expression for $h/r$,
\begin{align}
\frac{h}{r} = \left(\frac{\dot{M}'}{6\pi\alpha \rho_{\rm c} \Omega}\right)^{1/3} r^{-1}.
\end{align}
Once $h/r$ is estimated, Equation~\ref{eq:3} determines $T_{\rm c}$ from the assumed ideal equation of state. 

Using the disk solution, we construct a disk extending out to $\simeq10^{4}r_{\rm g}$ for $\Mbh=10^{6}\Msol$, corresponding to $2\times10^{4}\Rsol$, using $\simeq10^{7}$ cells. 
The disk parameters for our models are summarized in Table ~\ref{tab:initialparameter}.

\subsubsection{Stellar model}\label{subsubsec:star}
The initial state of the star in our hydrodynamics simulations was taken from stellar models evolved using the stellar evolution code {\sc MESA} (version r22.05.1) \citep{Paxton+2011,paxton:13,paxton:15,paxton:19,jermyn22}. We consider MS stars with three different masses, $\mstar=1\Msol$, $3\Msol$, and $10\Msol$, when the core H mass fraction is 0.3. The stellar radii of those stars are $R_{\star}=0.95\Rsol$, $2.5\Rsol$, and $4.7\Rsol$, respectively. Stars can grow in mass via accretion in the AGN disks \citep{Cantiello+2021}. The rate of accretion significantly influences the internal structure and chemical compositions of stars embedded in the disk. However, for those stars which approach the SMBH on a parabolic orbit from the effective radius of the nuclear cluster, $\simeq 0.5\pc\simeq 10^{7}r_{\rm g}$ for $\Mbh=10^{6}\Msol$ \citep{Neumayer+2020}, and are disrupted at the first pericenter passage, the accretion onto the star would not be significant. Assuming a Bondi–Hoyle accretion \citep{BondiHoyle1944,Bondi1952}\footnote{The Bondi radius $\propto 1/[c_{\rm s}^{2} + v^{2}]\propto v^{-2}$ where $v$ is the speed of the star because $c_{\rm s}\lesssim v$.} onto a $1\Msol$ star on a parabolic orbit in our disk with $\rho_{\rm c}=10^{-8}\gram\cm^{-3}$, the accretion rate can be estimated as $\dot{M}\simeq 10^{-15}\Msol \yr^{-1}(r/10^{7}r_{\rm g})^{-3/2}$. The total accreted mass until the star reaches the SMBH is $\simeq 10^{-10}\Msol$. Going one step further, because the dynamical friction is $\propto \dot{M}v$ \citep{Lee+2011,Lee+2014} where $v$ is the speed of the star, the total momentum of disk gas interacting with the star is many orders of magnitude smaller than the angular momentum of the stellar orbit.

We first map the 1D {\sc MESA} model into a 3D {\small AREPO} grid with $N\simeq 10^{6}$ cells using the mapping routine by \citet{Ohlmann+2017}. Then we fully relax the 3D single star, which usually takes up to five stellar dynamical times $\sqrt{R_{\star}^{3}/GM_{\star}}$. Figure~\ref{fig:relaxedstar} depicts the radial density profile of the fully relaxed stars in comparison with the MESA models. The internal profile of the 3D star matches the MESA model within less than a few \% except near the surface, corresponding to only a few $\%$ of the total mass, where the error is greater than 10\%.

\subsubsection{Initial conditions for star + disk model}\label{subsubsec:initial}

The relaxed stars are initially placed at $8r_{\rm t}$ on a parabolic orbit with a pericenter distance $r_{\rm p}\simeq 0.3 r_{\rm t}$ where $r_{\rm t}$ is the tidal disruption radius, defined as $r_{\rm t}\equiv (\Mbh/\mstar)^{1/3}\rstar$. The pericenter distance was chosen to ensure a complete disruption of the star in our fiducial model while keeping the events from becoming too relativistic.  We consider both prograde and retrograde orbits of the star relative to the orbit of the disk. Our fiducial models assume $\rho_{\rm c}=10^{-8}\gram\cm^{-3}$. In addition, we consider $\rho_{\rm c}=10^{-7}\gram \cm^{-3}$, $10^{-9}\gram \cm^{-3}$, $10^{-11}\gram \cm^{-3}$, and $10^{-12}\gram \cm^{-3}$. For reference, we perform the simulation of a TDE in an extremely low-density medium with  $\rho\simeq 10^{-20}\gram\cm^{-3}$, representing a vacuum, sharing the same encounter parameters of our fiducial model. To examine the stellar-mass dependence, we examine the post-disruption properties of the disk and debris for stars with $\mstar=3\Msol$ and $10\Msol$. For these cases, we only consider $\rho_{\rm c}=10^{-8}\gram \cm^{-3}$.

We performed convergence tests for the retrograde version of our fiducial models with different resolutions for the star ($N_{\star}$=125000, 250000, 500000, $10^{6}$ cells) and disk ($N_{\rm disk}=6\times10^{6}$ and $1.2\times 10^{7}$). By comparing several key quantities (e.g., debris mass as a function of radius from the black hole, the average radial mass infall rate towards the black hole), we confirmed that the results have already converged very well with $N_{\star}=250000$ and $N_{\rm disk}=6\times10^{6}$. To ensure the convergence, we chose $N_{\star}=10^{6}$. 

We summarize the model parameters in Table~\ref{tab:initialparameter}.

\begin{figure*}
    \centering
     \includegraphics[width=18cm]{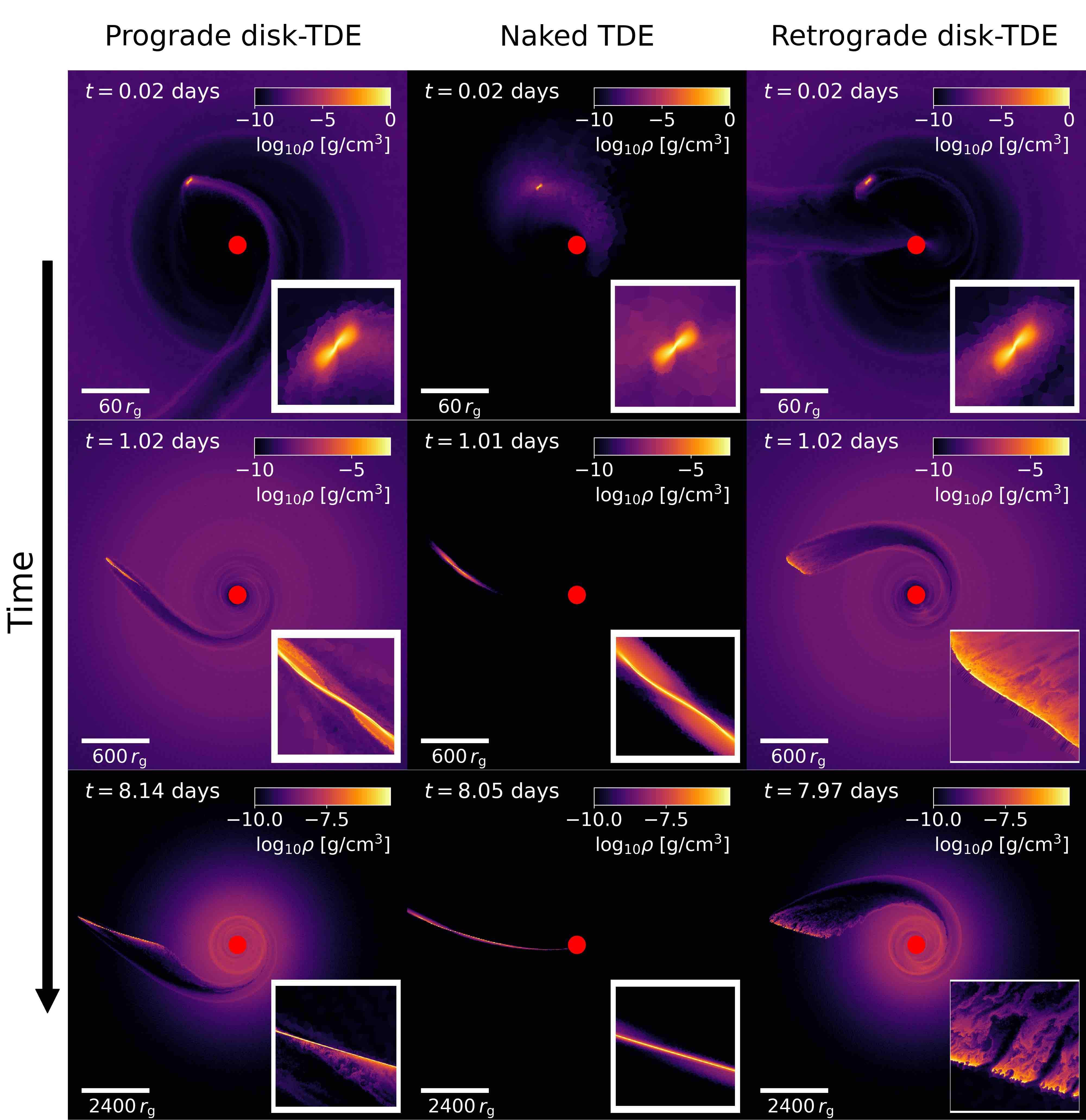}    
\caption{Successive moments in a full disruption event in a prograde (\textit{left}) and retrograde (\textit{right}) disk with a mid-plane density of $\rho_{\rm c}=10^{-8}$ g cm$^{-3}$. The \textit{middle} column shows them in a naked disruption event at the same times. The spatial scale shown in the inset corresponds to roughly 1/20 of the size indicated below the bar on the \textit{left-bottom} corner. Continuous interactions between the debris and the disk gas result in a significant perturbation of the debris's orbit and, therefore, its structure. The impact of the disk interaction is greater for the retrograde orbit than the prograde orbit. At later times, the debris is completely disintegrated and mixed into the disk.}
	\label{fig:example}
\end{figure*}

\begin{figure*}
    \centering
    \includegraphics[width=8.4cm]{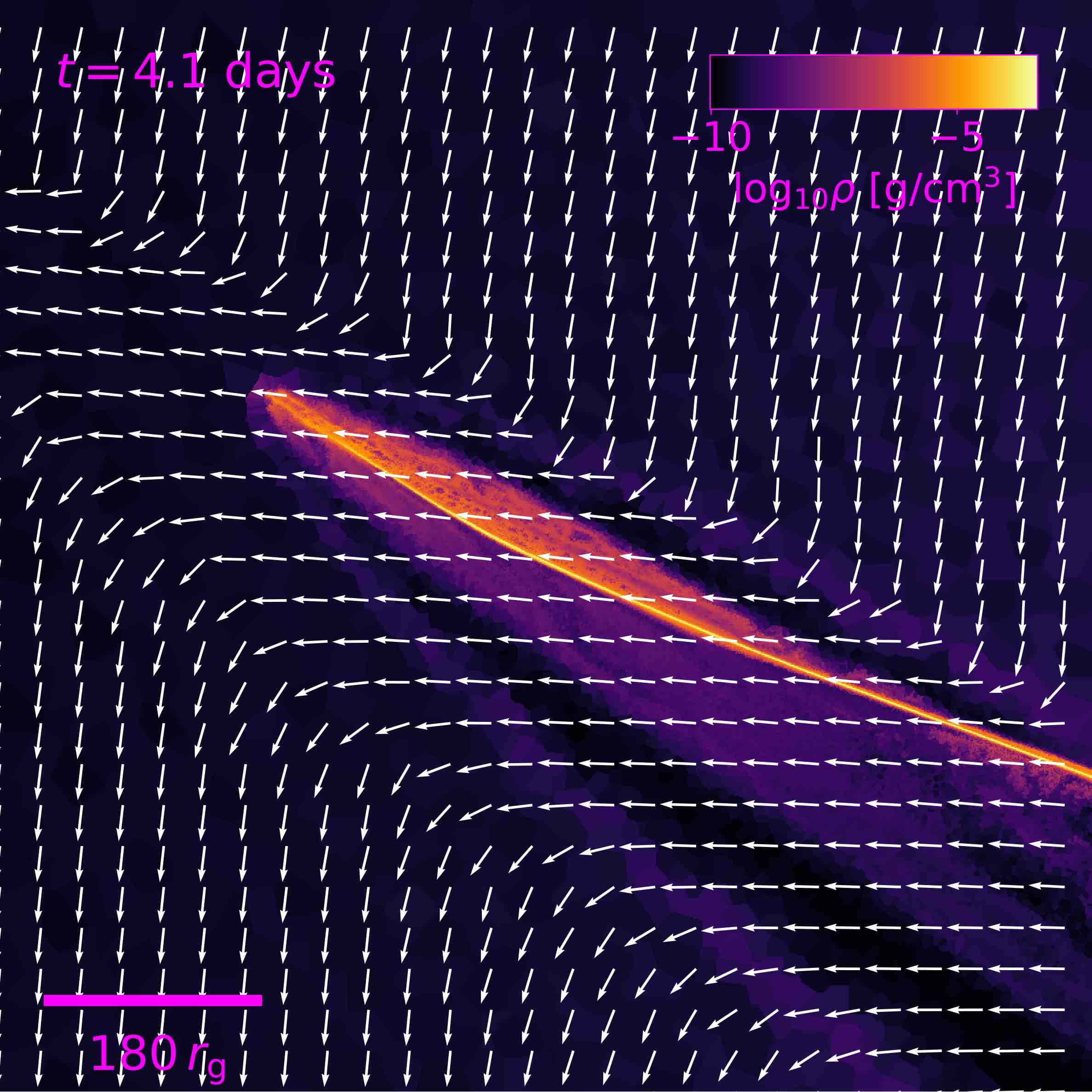}
    \includegraphics[width=8.4cm]{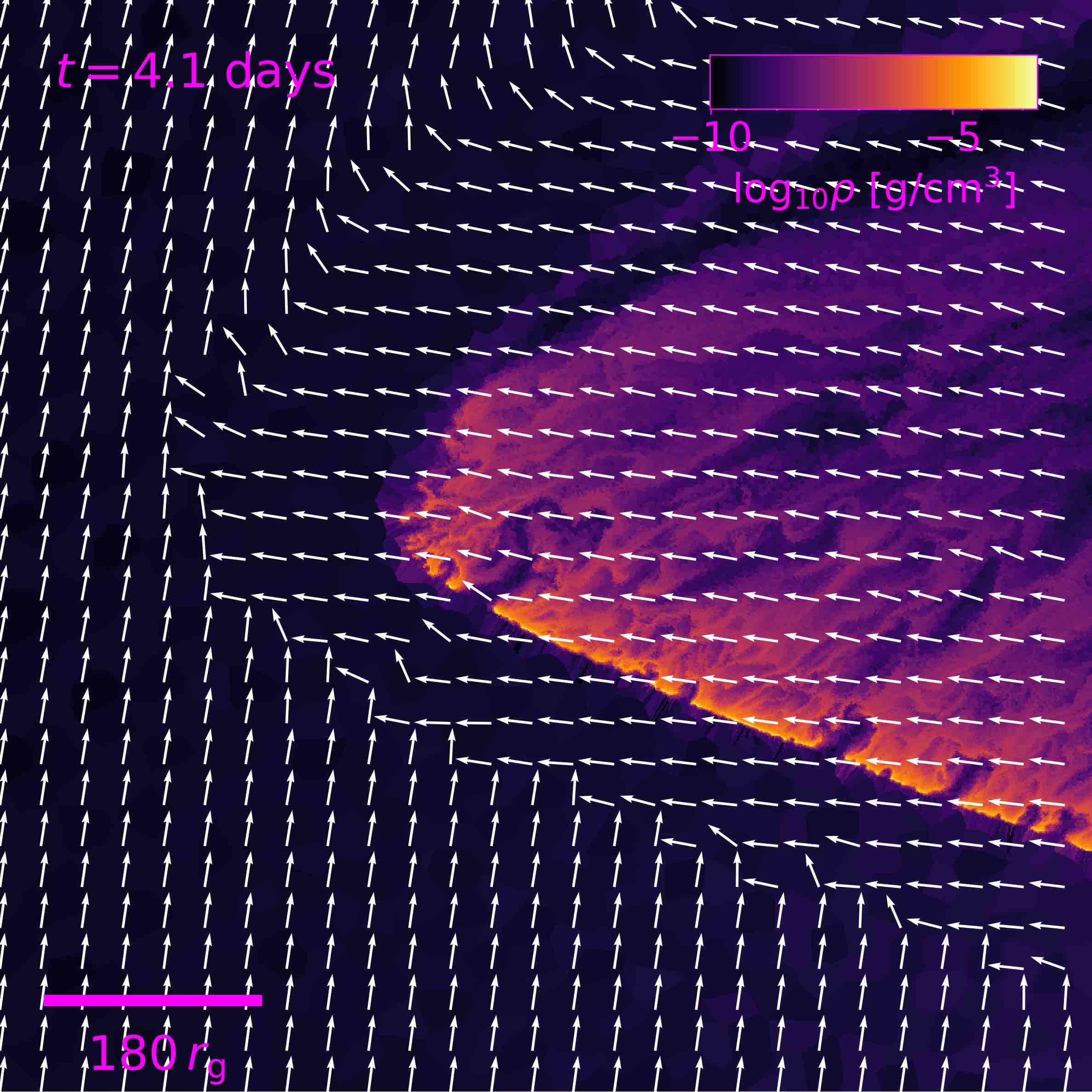}
\caption{Zoom-in near the head of the debris for the prograde (\textit{left}) and retrograde (\textit{right}) cases with $\rho_{\rm c}=10^{-8}\gram \cm^{-3}$, measured at $\sim4$ days after disruption. The white arrows show the direction of motion of the gas. In the prograde case, the disk interactions act to increase the angular momentum of the debris whereas, in the retrograde case, the disk interactions cancel out the angular momentum of the debris.}
	\label{fig:flowmotion}
\end{figure*}

\section{Results}\label{sec:results}

\subsection{Overview}

We first provide a qualitative overview of the results of our simulations. More quantitative descriptions will be given in the following sections. 

Figure~\ref{fig:example} shows successive moments in a full disruption of the $1\Msol$ star in our fiducial models ($\rho_{\rm c}=10^{-8}\gram \cm^{-3}$) with a prograde (\textit{left}) and retrograde (\textit{right}) orbit. For comparison, we show in the \textit{middle} column the same moments in a naked TDE sharing the same encounter parameters. The pre-disruption orbit and the internal structure of the star are not significantly affected by the disk provided a negligible amount of disk mass is interacting with the star until it reaches pericenter (\textit{$1^{\rm st}$} panels). Upon disruption, the debris starts to expand in size. The increasingly larger cross-sections makes the debris more subject to interacting with the disk  (\textit{$2^{\rm nd}-3^{\rm rd}$} panels). Depending on whether the orbit is prograde or retrograde, the evolution of the debris can be qualitatively different, meaning potentially different observational signatures. 

In the prograde case, the outer edges of the debris are gradually mixed into the disk via the Rayleigh–Taylor instability \citep{Taylor+1950,Rayleigh}. Due to the coherent motion between the debris and disk, as illustrated in the \textit{left} panel of Figure~\ref{fig:flowmotion}, the interaction with the disk acts to add angular momentum to the debris. On the other hand, the evolution of debris in the retrograde case is more dramatic due to the significant cancellation of angular momentum. In the retrograde case, like the prograde case, the debris is continuously lost to the disk. But the mixing is more violent, which is shown in the \textit{$3^{\rm rd}$} \textit{right} panel of Figure~\ref{fig:flowmotion}. As a result, the initially coherent motion of the debris is significantly perturbed even before any of the bound matter starts to return to the SMBH. Because of increasingly irregular perturbations caused by the Rayleigh–Taylor instability, the energy distribution and the resulting fallback curve of the debris tend to be bumpier than that for the prograde case. In the case of a sufficiently high mid-plane density (e.g., $\rho_{\rm c}=10^{-8} \gram \cm^{-3}$), the entire debris can be mixed with the disk in less time than the peak fallback time of debris in a naked TDE with the same disruption parameters.

\begin{figure*}
    \centering
    \includegraphics[width=8.6cm]{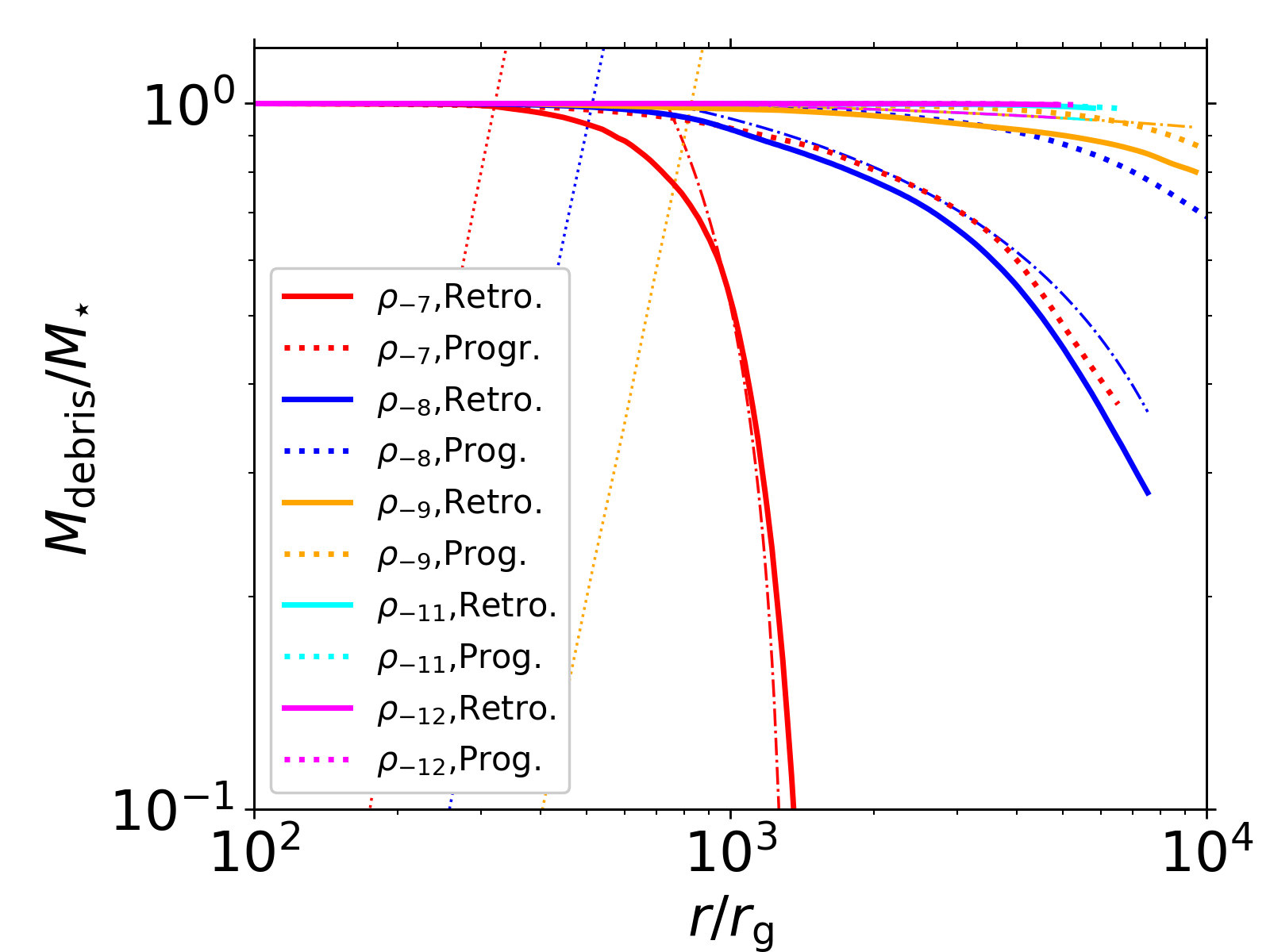}
    \includegraphics[width=8.6cm]{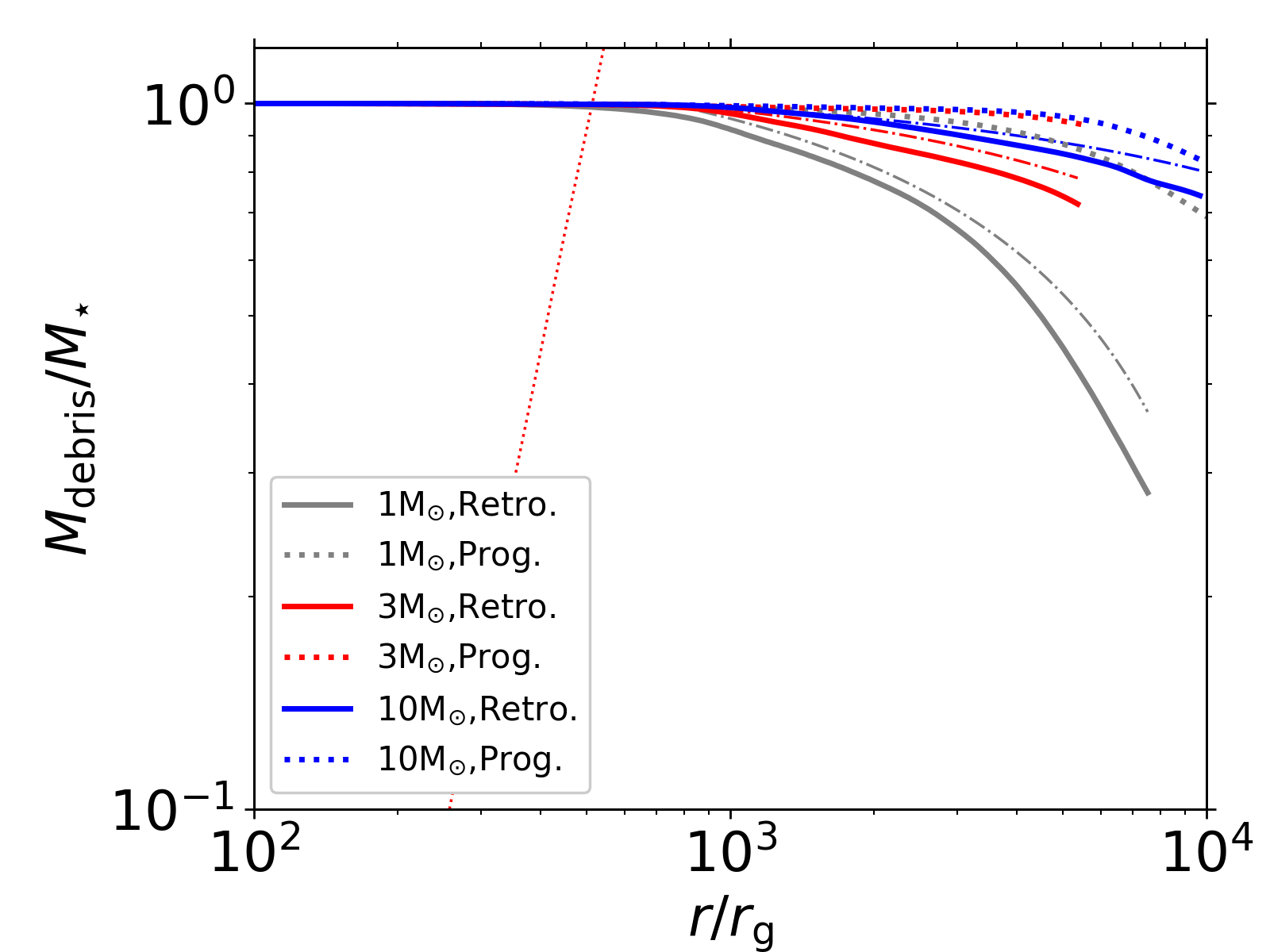}
\caption{Time evolution of the fractional remaining debris mass that has not been mixed into the disk as a function of the distance of the center of mass of the debris from the black hole for different models: (\textit{left}) different disk densities and (\textit{right}) different stellar masses. The solid (dotted) lines represent the prograde (retrograde) cases. The dot-dashed diagonal lines in both panels indicate the prediction from our semi-analytic model for the retrograde cases (\S\ref{subsec:massloss}). The dotted diagonal lines show the enclosed mass of the disk: in the \textit{left} panel, the colors of the lines for the disk enclosed mass match those for the debris mass while in the \textit{right} panel, the disk mass is for $\rho_{\rm c}=10^{-8}$ g cm$^{-3}$. }
	\label{fig:debris_mass}
\end{figure*}

\subsection{Debris mass loss - semi-analytic approach}\label{subsec:massloss}

To quantify the mass loss of debris to the disk, we distinguish the debris from the disk gas using a passive scalar. The passive scalar is an artificial scalar quantity initially assigned to each cell. The scalar then evolves via advection. The initial value of the passive scalar for the cells that belong to the stars is one, and for the disk cells, it is zero. Depending on the mass exchange (and thus momentum exchange) between the cells, the passive scalar varies between zero (for disk cells) and one (for cells originally in the stars). Identifying a specific region of gas with a passive scalar has been used in the literature to investigate mixing of gas in various contexts \citep[e.g.,][]{McCourt+2015,Gronke+2018,Dutta+2019,Kanjilal+2021,Farber+2022,Farber+2022b}.Our close investigation of the distribution of the scalar suggests that the scalar quantity for the debris in a coherent motion is generally larger than 0.99, meaning it mixes with the disk material by roughly less than 1\% in mass.

Figure~\ref{fig:debris_mass} illustrates the fractional mass of debris relative to the initial stellar mass as a function of the distance of the center of mass of debris from the SMBH. The \textit{left} (\textit{right}) panel compares the debris mass between models with different mid-plane disk densities (stellar masses). The most noticeable difference is between the prograde and retrograde cases. For the case with the retrograde orbit relative to the disk with the highest density ($\rho_{\rm c}=10^{-7}\gram \cm^{-3}$, red solid line in the \textit{left} panel), the entire debris gets mixed to the disk near the density cusp at $r\simeq 10^{3} r_{\rm g}$. On the other hand, for the prograde case with the same disk, the mass loss is less severe: $\simeq 30\%$ of the debris survives until it reaches 7000$r_{\rm g}$. As the disk mid-plane density decreases or the stellar mass increases, a larger fraction of debris can reach farther out.

The impact of the debris mass loss on the disk structure would be insignificant because the enclosed mass of the disk (dotted diagonal lines in Figure~\ref{fig:debris_mass}) is many orders of magnitude greater than the stellar mass by the time the entire debris is dissociated and mixed into the disk.

We can understand the trends of debris mass loss to the disk by comparing how much disk mass interacts to remove the momentum of the debris along the way out. To this end, we build a semi-analytic model for the mass of debris that is mixed to the disk  in the retrograde case, which allows us to estimate the maximum distance that the debris can travel through a disk. For the prograde case, the momentum of the disk is added to the debris (see Figure~\ref{fig:dlde}), so this semi-analytic model won't apply to the prograde case.

We assume that the disruption of the coherent motion of the debris is primarily governed by the amount of mass of the disk flow hitting the debris. In other words, the remaining debris mass $M_{\rm debris}$ that continues to follow the orbit that the debris would have assuming a ballistic orbit is simply the initial debris mass or stellar mass $M_{\star}$ minus the mass of disk flow  $M_{\rm d}$ continuously interacting with the debris,
\begin{align}
    M_{\rm debris}(t) = M_{\star} - M_{\rm d}(t).
\end{align}
When the self-gravity is not important, the slow-down of the debris would naturally lead to mixing into the disk. However, when the self-gravity is strong, the entire debris would slow down instead of mixing into the disk. The former is more relevant for AGN-TDEs.

We may be able to estimate $M_{\rm d}(t)$ as,
\begin{align}
    M_{\rm d}(t)\sim \int_{0}^{t}\rho_{\rm disk}(r) v_{\rm disk}(r) A_{\rm debris}(t') dt',
\end{align}
where $\rho_{\rm disk}(r)$ is the density at distance $r$ from the SMBH, $v_{\rm disk}(r)\simeq \sqrt{G\Mbh/r}$ the flow speed at $r$, and $A_{\rm debris}$ the cross-section of the debris whose normal is parallel to the disk flow. Although each part of the debris moves at a different speed, we simply assume that the entire debris continues to follow the original orbit of the star, i.e., parabolic orbit, so that $r(t)$ since disruption is expressed,
\begin{align}\label{eq:rpara}
    r(t) = \left(\frac{9 G \Mbh}{2}\right)^{1/3} t^{2/3},
\end{align}
and the radial velocity is,
\begin{align}\label{eq:vrpara}
    v_{\rm debris}(t) = \frac{dr(t)}{dt} = \frac{2}{3}\left(\frac{9 G \Mbh}{2}\right)^{1/3} t^{-1/3}.
\end{align}
With Equations \ref{eq:rpara} and \ref{eq:vrpara}, $\rho_{\rm disk}(r)$ and $v_{\rm disk}(r)$ are now a function of time. To calculate $A_{\rm debris}(t)$, we assume that by the time the debris arrives at $r\simeq r_{\rm t}$ since disruption, the debris extends to $l_{\rm debris}\simeq \alpha \rstar$ with a width $w\simeq \rstar$, where $\alpha\simeq 20 - 22$ from our simulations (see Figure~\ref{fig:example}). We further assume that the debris expands in size such that $l\propto t^{4/3}$ and $w_{\rm debris}\propto t ^{1/3}$ before the most bound debris starts to return \citep{Coughlin2016,BonnerotStone2021}. These assumptions allow us to write an expression for $l_{\rm debris}$ and $w_{\rm debris}$,
\begin{align}\label{eq:l}
     l_{\rm debris} &\simeq \alpha \rstar \left(\frac{t} {t(r=r_{\rm t})}\right)^{4/3},\nonumber\\
                    &\simeq \alpha \Rsol \left(\frac{\mstar}{1\Msol}\right)^{2/3}\left(\frac{\rstar}{1\Rsol}\right)^{-1}\left(\frac{t}{0.01{\rm days}}\right)^{4/3},
\end{align}
and
\begin{align}\label{eq:h}
     w_{\rm debris} &\simeq \rstar \left(\frac{t} {t(r=r_{\rm t})}\right)^{1/3},\nonumber\\
                    &\simeq \Rsol \left(\frac{\mstar}{1\Msol}\right)^{1/6}\left(\frac{\rstar}{1\Rsol}\right)^{1/2}\left(\frac{t}{0.01{\rm days}}\right)^{1/3},
\end{align}
where $t(r=r_{\rm t})$ is estimated using Equation~\ref{eq:rpara}. Note that the average density of debris $\bar{\rho}_{\rm debris} \propto M_{\star}/[l_{\rm debris}h_{\rm debris}^{2}]\propto t^{-2}$, which we have confirmed from our simulations. It follows that the cross-section $A_{\rm debris}$ is,
\begin{align}
    A_{\rm debris} \simeq l_{\rm debris}w_{\rm debris}\simeq \alpha \Rsol^{2} \left(\frac{M_{\star}}{1\Msol}\right)^{5/6}\left(\frac{\rstar}{1\Rsol}\right)^{-1/2}\left(\frac{t}{0.01{\rm days}}\right)^{5/3}.
\end{align}
Because the disk density profile has two regions, i.e., flat for $r<r_{\rm cusp}=10^{3}r_{\rm g}$ and power-law for $r>r_{\rm cusp}$, we will calculate the mass loss due to disk-debris interaction for the two regions separately. 
\begin{enumerate}
    \item \textit{Flat region ($\rho_{\rm disk}=\rho_{\rm c}$}): the time required for the debris to reach the cusp is roughly estimated using Equation \ref{eq:rpara},
    \begin{align}
        t_{\rm cusp} = t(r=r_{\rm cusp})\simeq 0.85 {\rm days} \left(\frac{r_{\rm cusp}}{10^{3}r_{\rm g}}\right)^{3/2}\left(\frac{\Mbh}{10^{6}\Msol}\right)^{-1/2}.
    \end{align}
    So for $t<t_{\rm cusp}$, the mass loss is,
    \begin{align}\label{eq:dm1}
        M_{\rm d}^{r<r_{\rm cusp}}(t)&\simeq \int_{0}^{t}\rho_{\rm c} v_{\rm disk}(r(t')) A_{\rm debris}(t') dt',\nonumber\\
    &=0.05\Msol \left(\frac{\alpha}{22}\right) \left(\frac{\Mbh}{10^{6}\Msol}\right)^{1/3}\left(\frac{M_{\star}}{1\Msol}\right)^{5/6}\left(\frac{\rstar}{1\Rsol}\right)^{-1/2}\nonumber\\
    &\times\left(\frac{\rho_{\rm c}}{10^{-8}\gram\cm^{-3}}\right)\left(\frac{t}{0.85{\rm days}}\right)^{7/3}.
    \end{align}
    As shown in the equation, the fractional mass loss $M_{\rm d}^{r<r_{\rm cusp}}/M_{\star}$ has a relatively weak dependence on $M_{\star}$ ($\propto M_{\star}^{-1/6}$) and $\Mbh$ ($\propto \Mbh^{1/3}$), but rather strongly depends on $\rho_{\rm c}$. For example, roughly half the debris mass would be mixed to the disk at $r\simeq r_{\rm cusp}$ (or $t\simeq 0.85$ days) when $\rho_{\rm c}\simeq 10^{-7} \gram \cm^{-3}$, which is illustrated in the \textit{left} panel (red solid line) of Figure~\ref{fig:debris_mass}.

\begin{figure*}
    \centering
    \includegraphics[width=5.8cm,trim={0.9cm 0 0 0},clip]{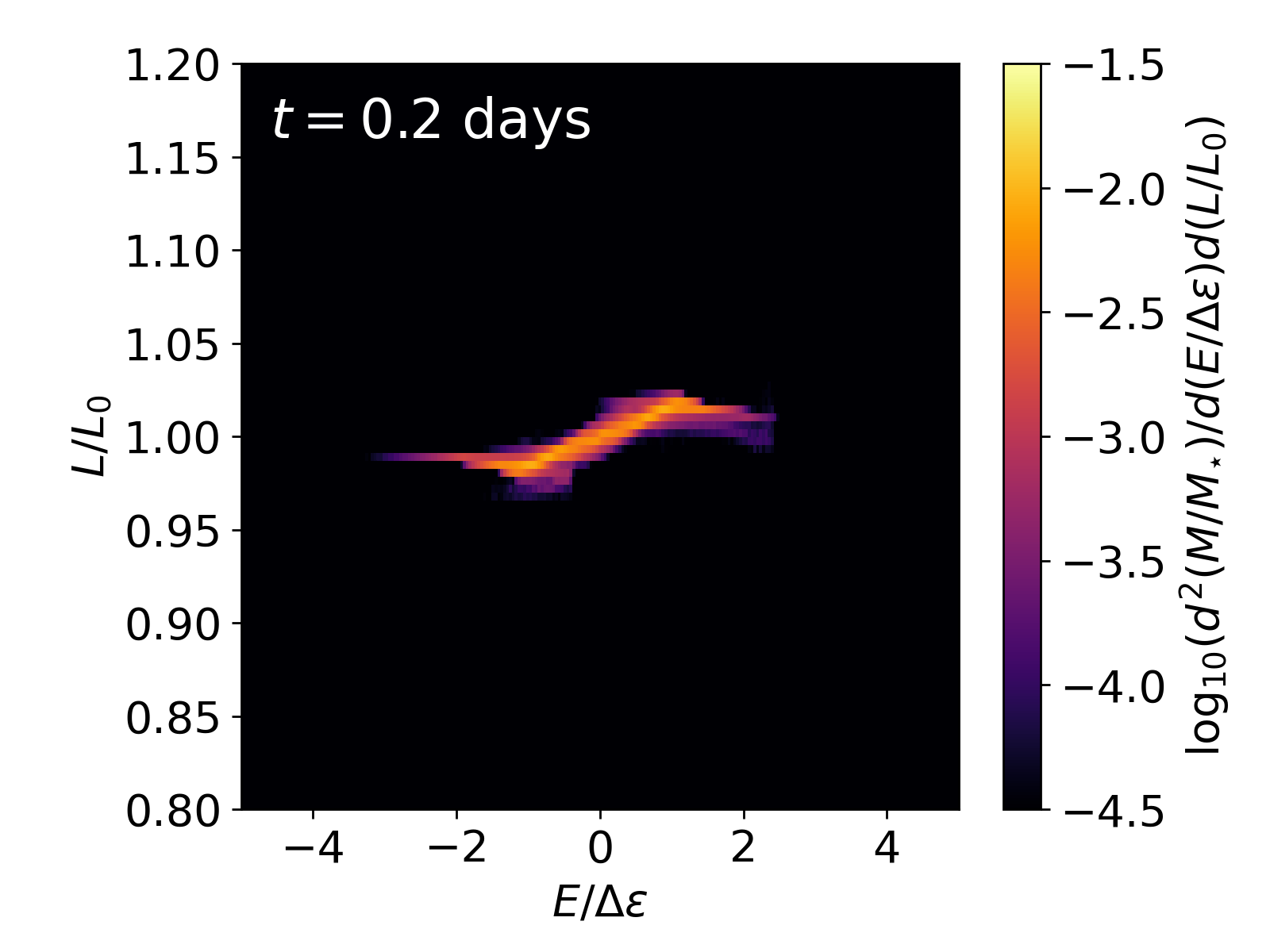}
    \includegraphics[width=5.8cm,trim={0.9cm 0 0 0},clip]{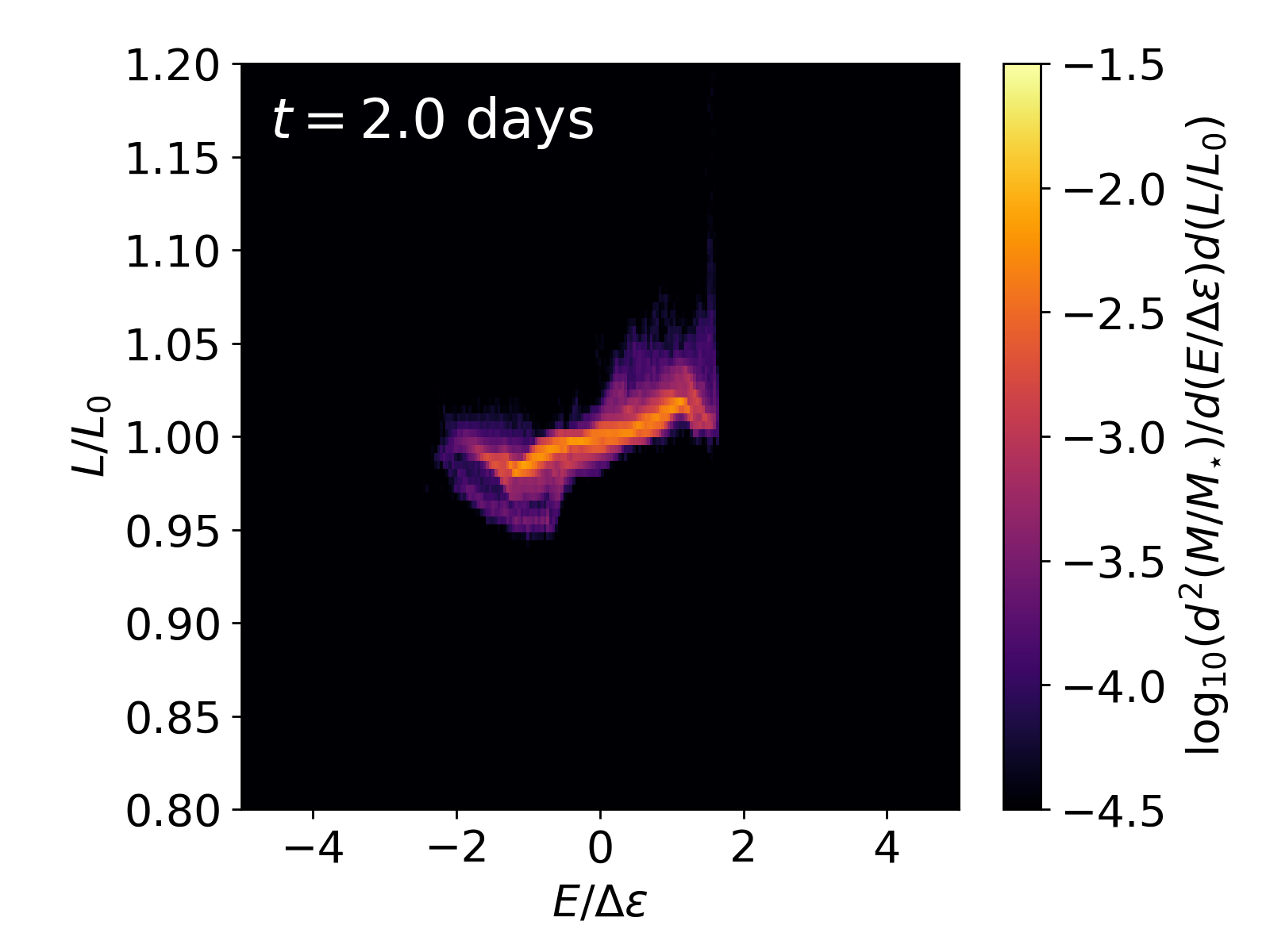}
    \includegraphics[width=5.8cm,trim={0.9cm 0 0 0},clip]{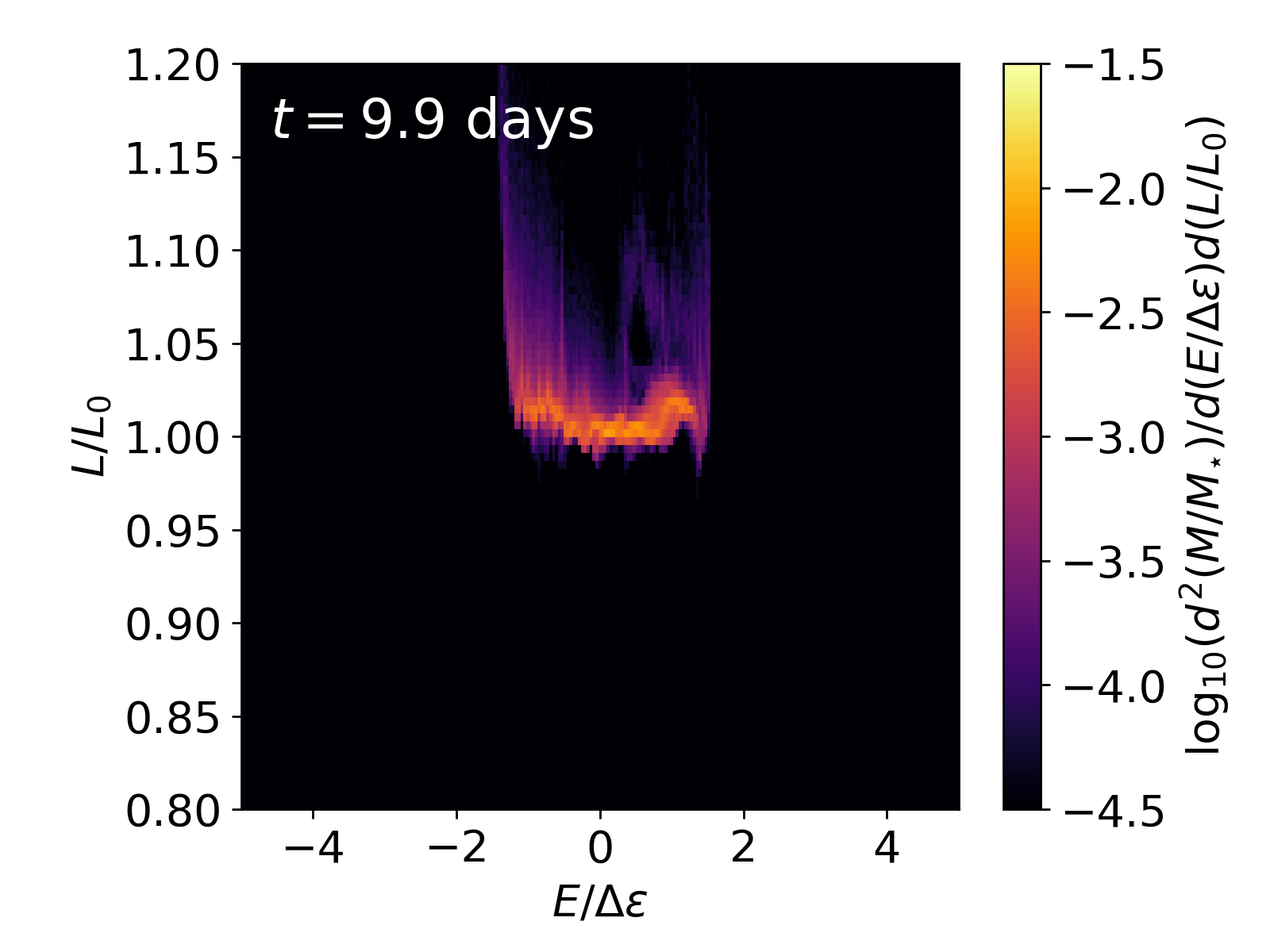}\\
    \includegraphics[width=5.8cm,trim={0.9cm 0 0 0},clip]{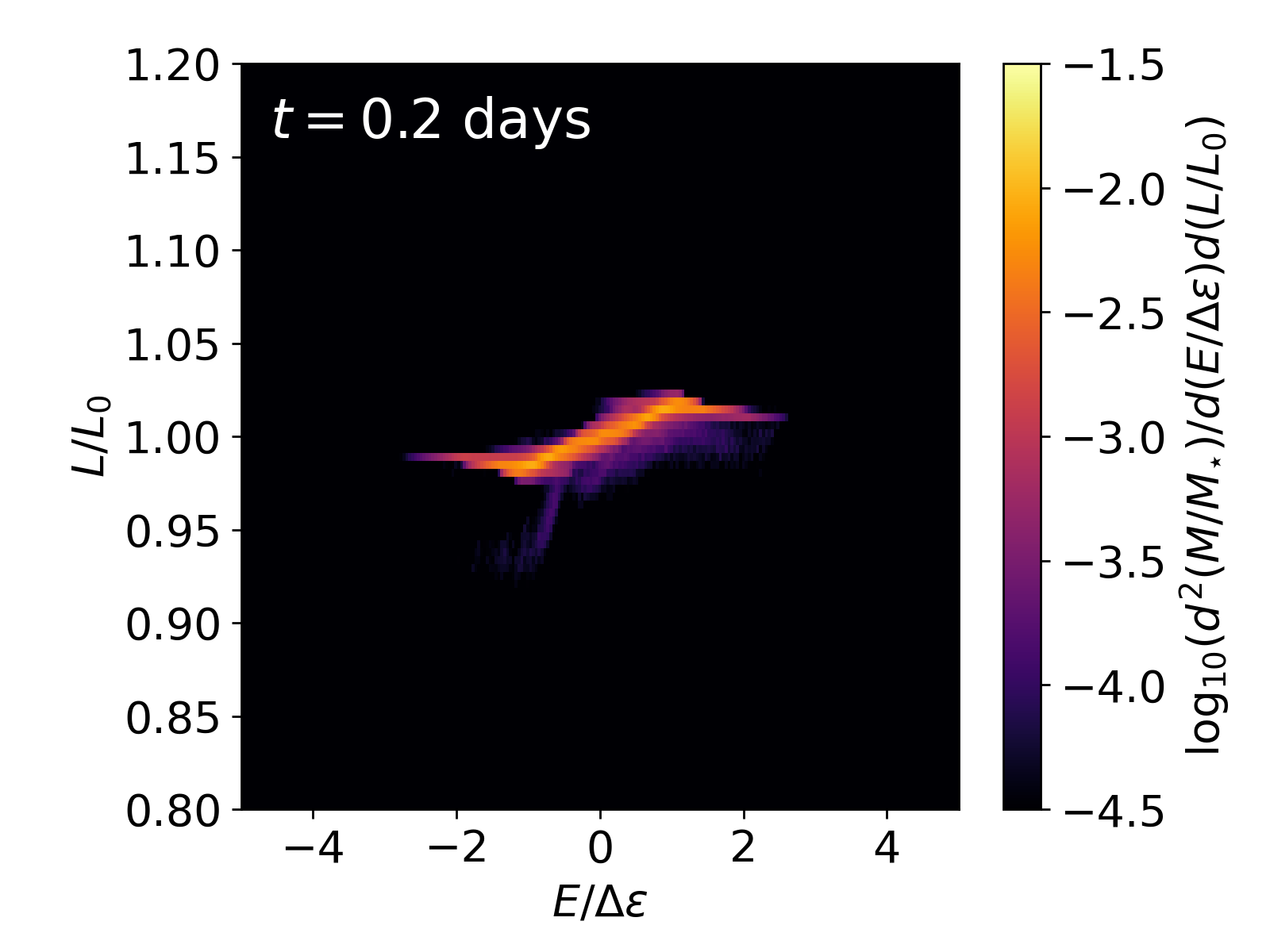}
    \includegraphics[width=5.8cm,trim={0.9cm 0 0 0},clip]{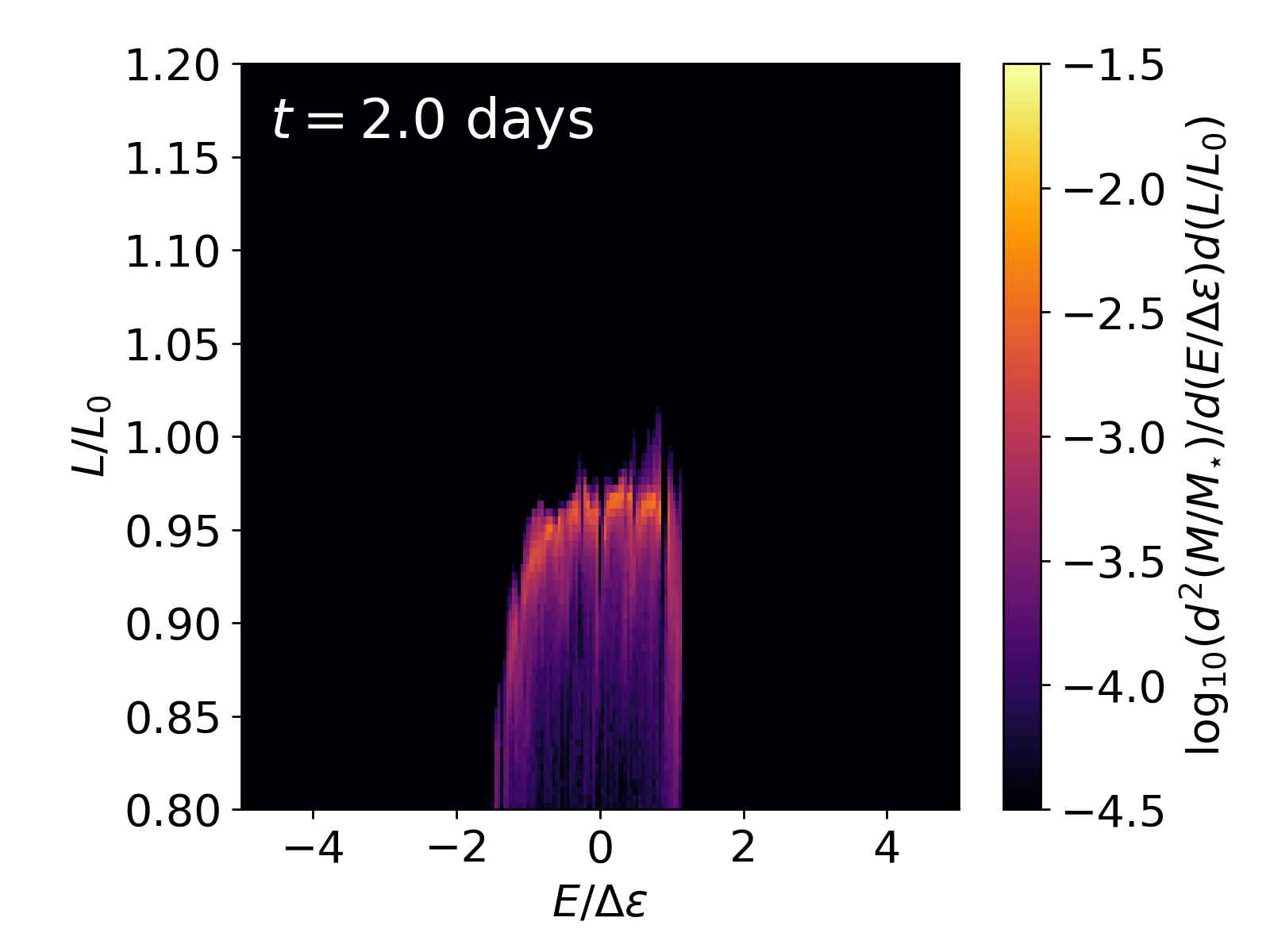}
    \includegraphics[width=5.8cm,trim={0.9cm 0 0 0},clip]{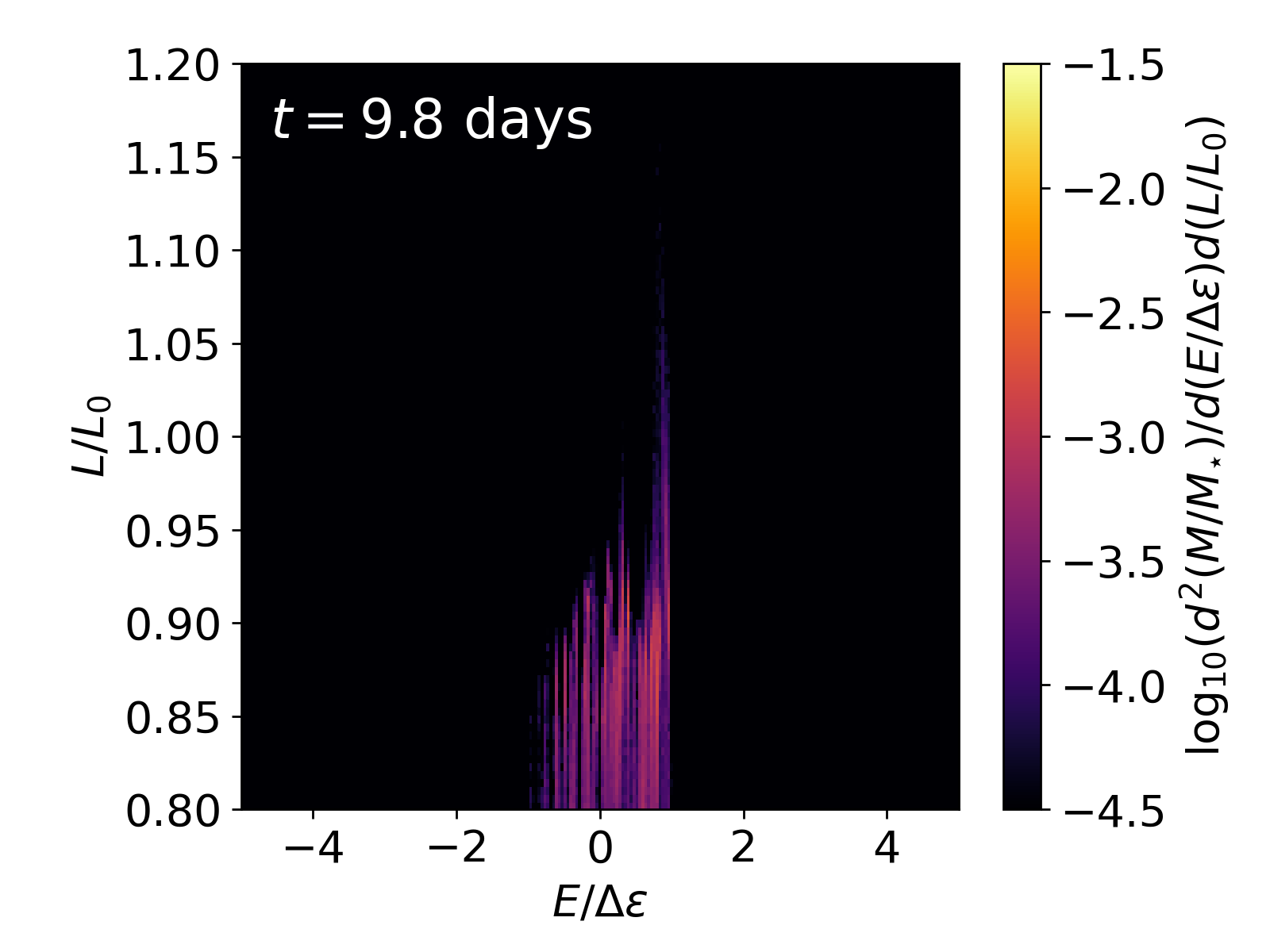}
\caption{Distribution of specific energy $E$ and specific angular momentum $L$ for a full disruption of a $1\Msol$ star on a prograde (\textit{top}) and retrograde (\textit{bottom}) orbit relative to that of a disk with $\rho_{\rm c}=10^{-8}\gram\cm^{-3}$ around a $10^{6}\Msol$ SMBH at three different times. }
	\label{fig:dlde}
\end{figure*}
    
    \item\textit{Power-law region ($\rho_{\rm disk}\propto r^{p}$ with $p=-3$)}: the mass loss at $ t_{\rm cusp}\lesssim t \lesssim t_{0}$ is,
    \begin{align}\label{eq:dm2}
        &M_{\rm d}^{r>r_{\rm cusp}}(t)\nonumber\\
        &\simeq M_{\rm d}^{r<r_{\rm cusp}}(t=t_{\rm cusp}) + \int_{t_{\rm cusp}}^{t}\rho_{\rm c} \left(\frac{r(t')}{r_{\rm cusp}}\right)^{-3} v_{\rm disk}(r(t')) A_{\rm debris}(t') dt',\nonumber\\
    &=\left(\frac{\alpha}{22}\right) \left(\frac{\Mbh}{10^{6}\Msol}\right)^{-2/3}\left(\frac{M_{\star}}{1\Msol}\right)^{5/6}\left(\frac{\rstar}{1\Rsol}\right)^{-1/2}\left(\frac{r_{\rm cusp}}{10^{3}r_{\rm g}}\right)^{3}\left(\frac{\rho_{\rm c}}{10^{-8}\gram\cm^{-3}}\right)\nonumber\\
    &\times\left[-0.3\Msol\left(\frac{r_{\rm cusp}}{10^{3}r_{\rm g}}\right)^{1/2}\left(\frac{\Mbh}{10^{6}\Msol}\right)^{-1/6} + 0.85\Msol \left(\frac{t}{15{\rm days}}\right)^{7/3}\right].
    \end{align}
    The mass loss predicted from the semi-analytic model is in good agreement with the simulation results for the retrograde cases (solid lines), which are depicted in Figure~\ref{fig:debris_mass} using dot-dashed lines. Our semi-analytic model suggests that in our fiducial model with $\rho_{\rm c}=10^{-8}\gram \cm^{-3}$, roughly 50\% (80\%) of the debris mass would be mixed into the disk in 15 days (30 days $\simeq t_{0}$) while the debris is moving away from the SMBH. Among the rest of the remaining debris (20\%), the bound part would have to plow through the disk inwards while it was returning to the SMBH, such that it is very likely that at least the remaining bound debris would be completely mixed into the disk on the way in. In fact, we do not observe any coherent return of debris to the SMBH in our simulations. This means, no TDE-like flare would be generated. 
\end{enumerate}

\begin{figure*}
    \centering
    \includegraphics[width=8.6cm]{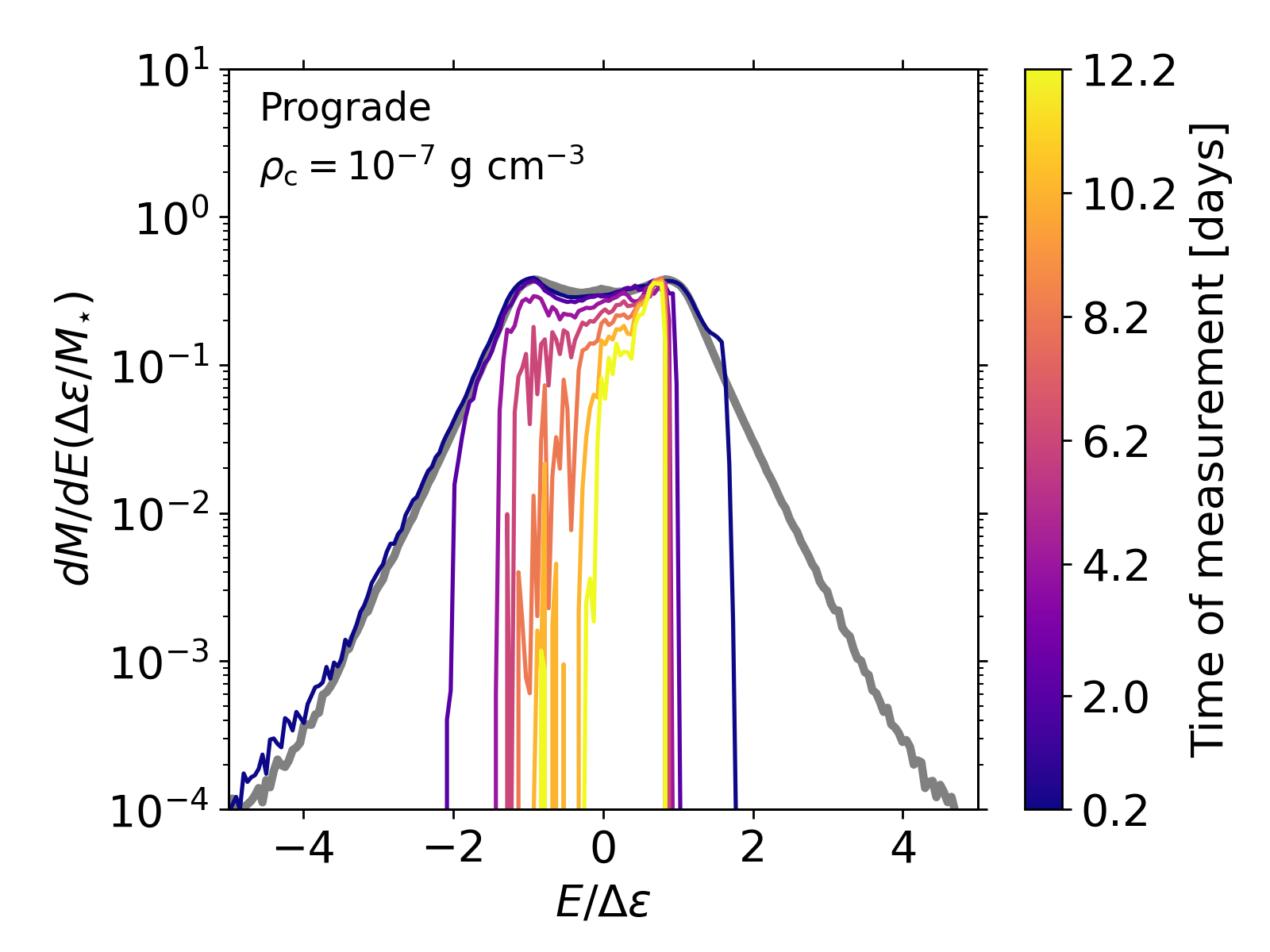}
    \includegraphics[width=8.6cm]{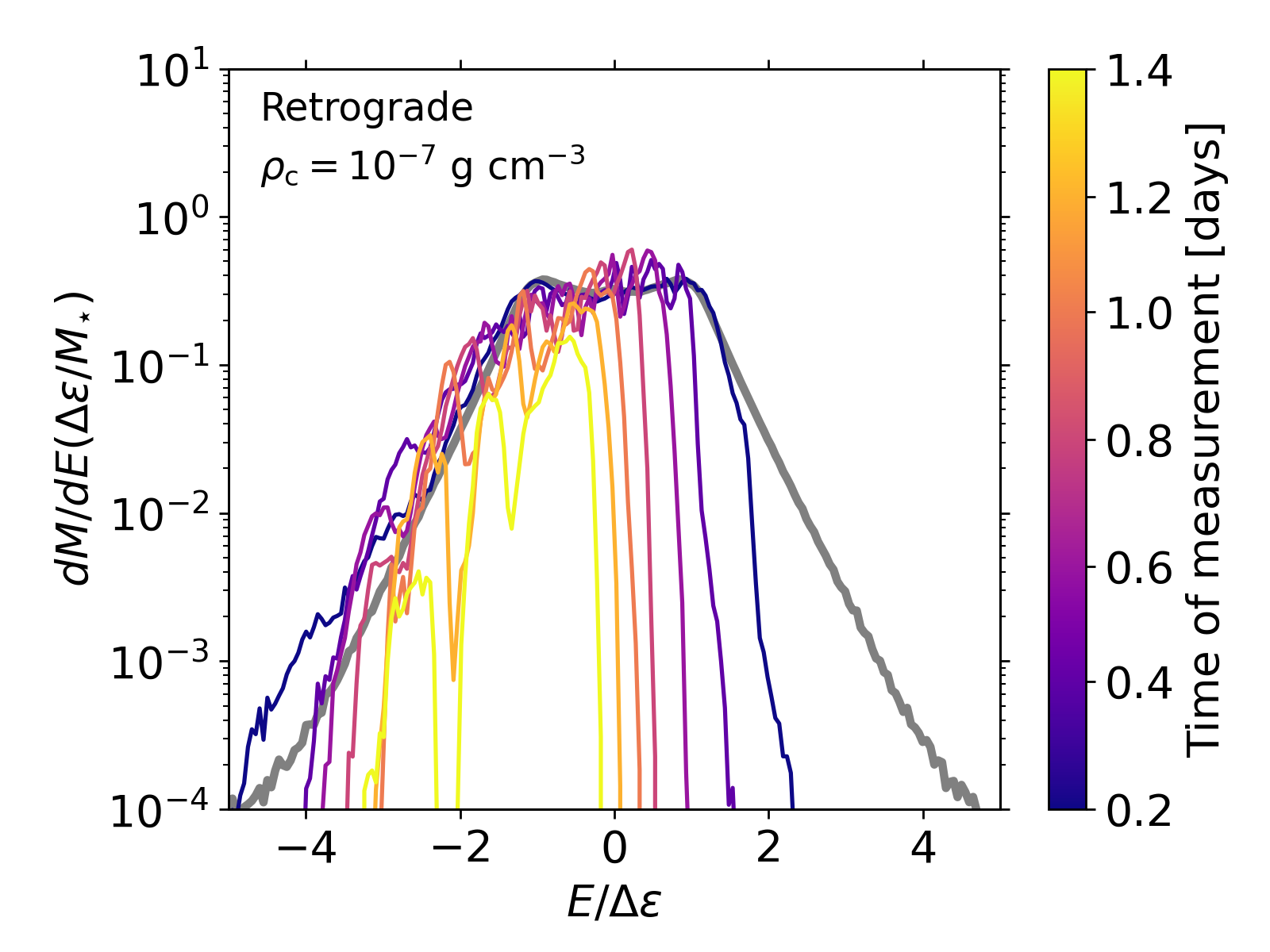}\\
    \includegraphics[width=8.6cm]{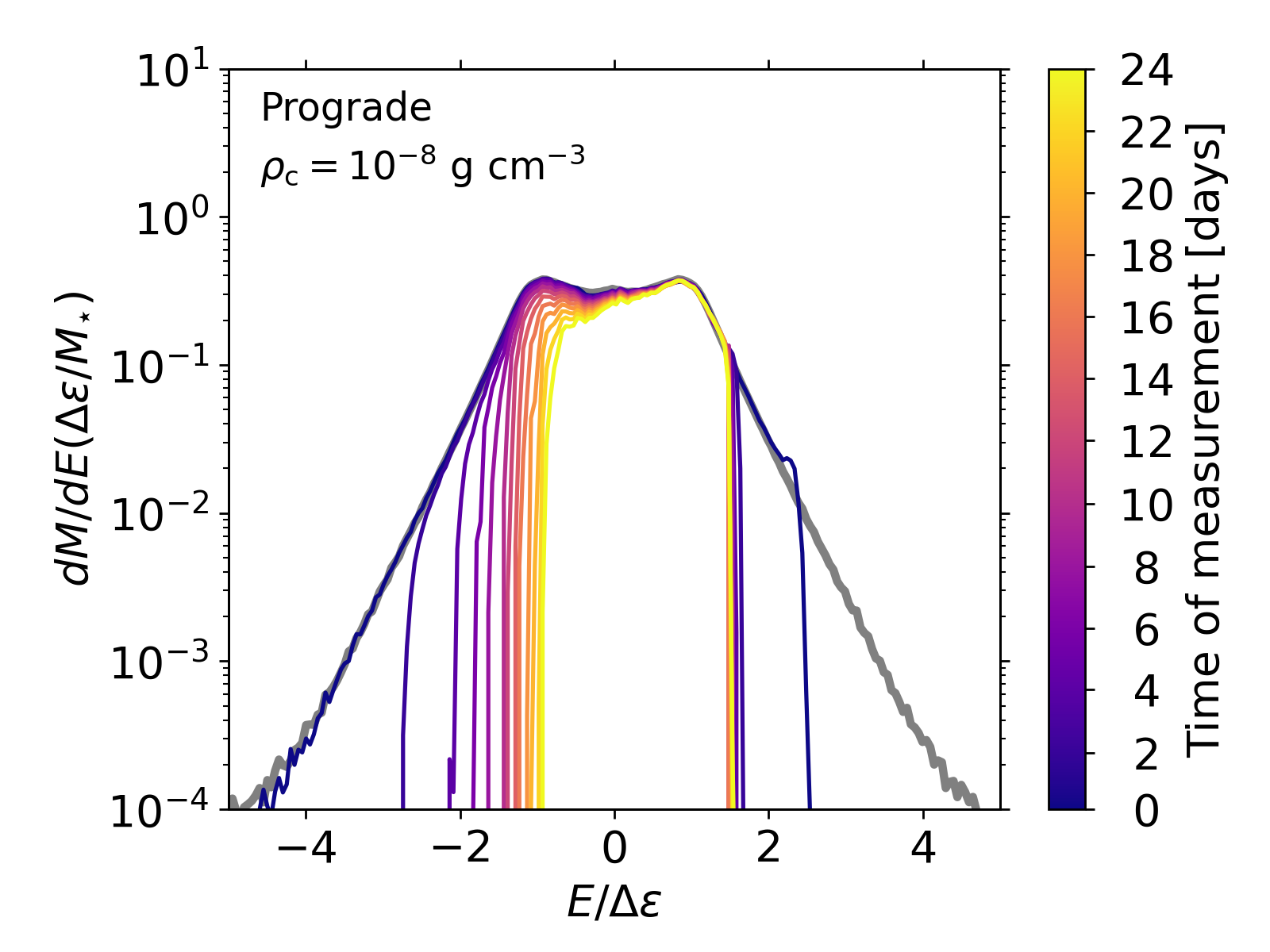}
    \includegraphics[width=8.6cm]{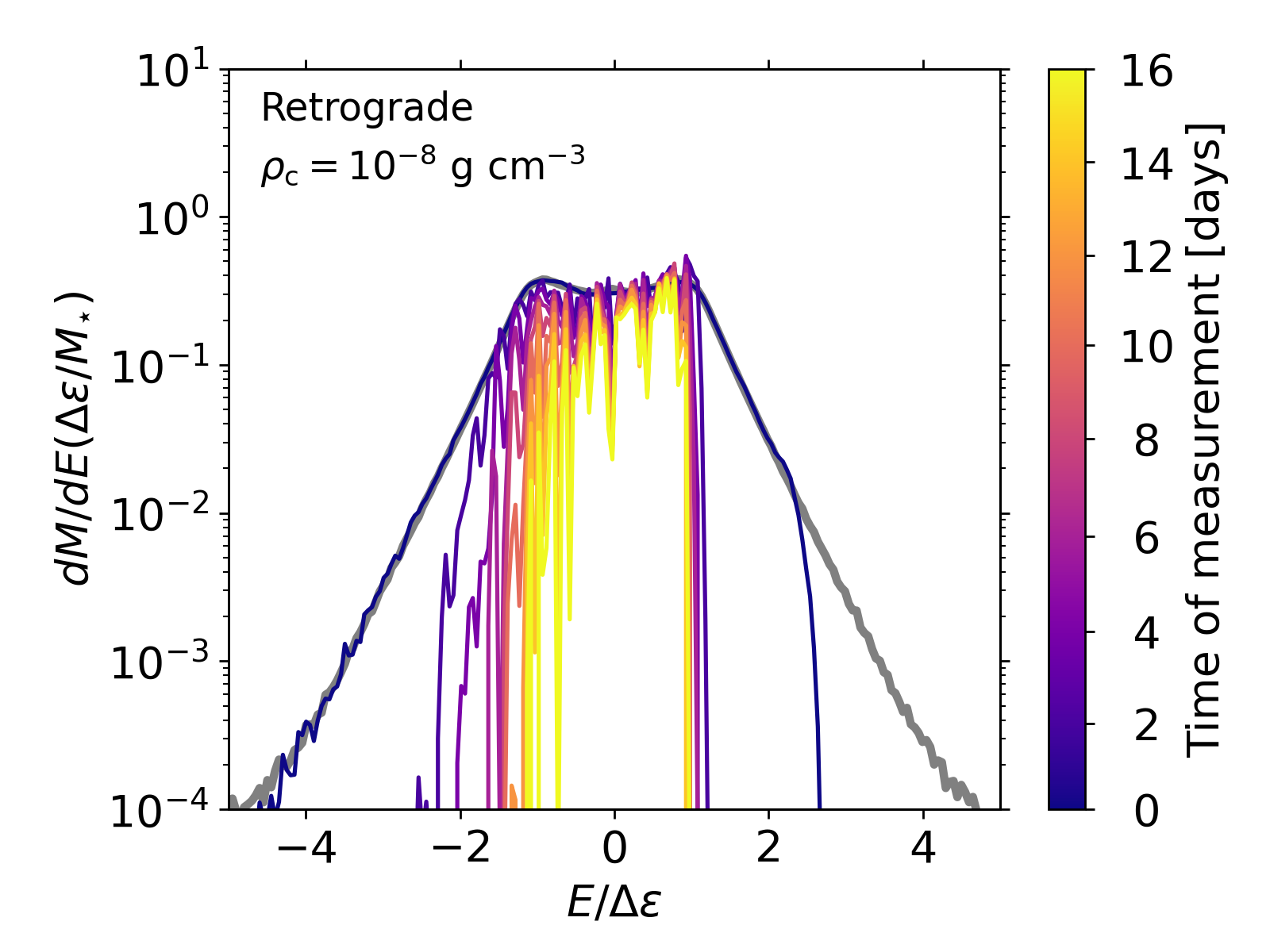}\\
    \includegraphics[width=8.6cm]{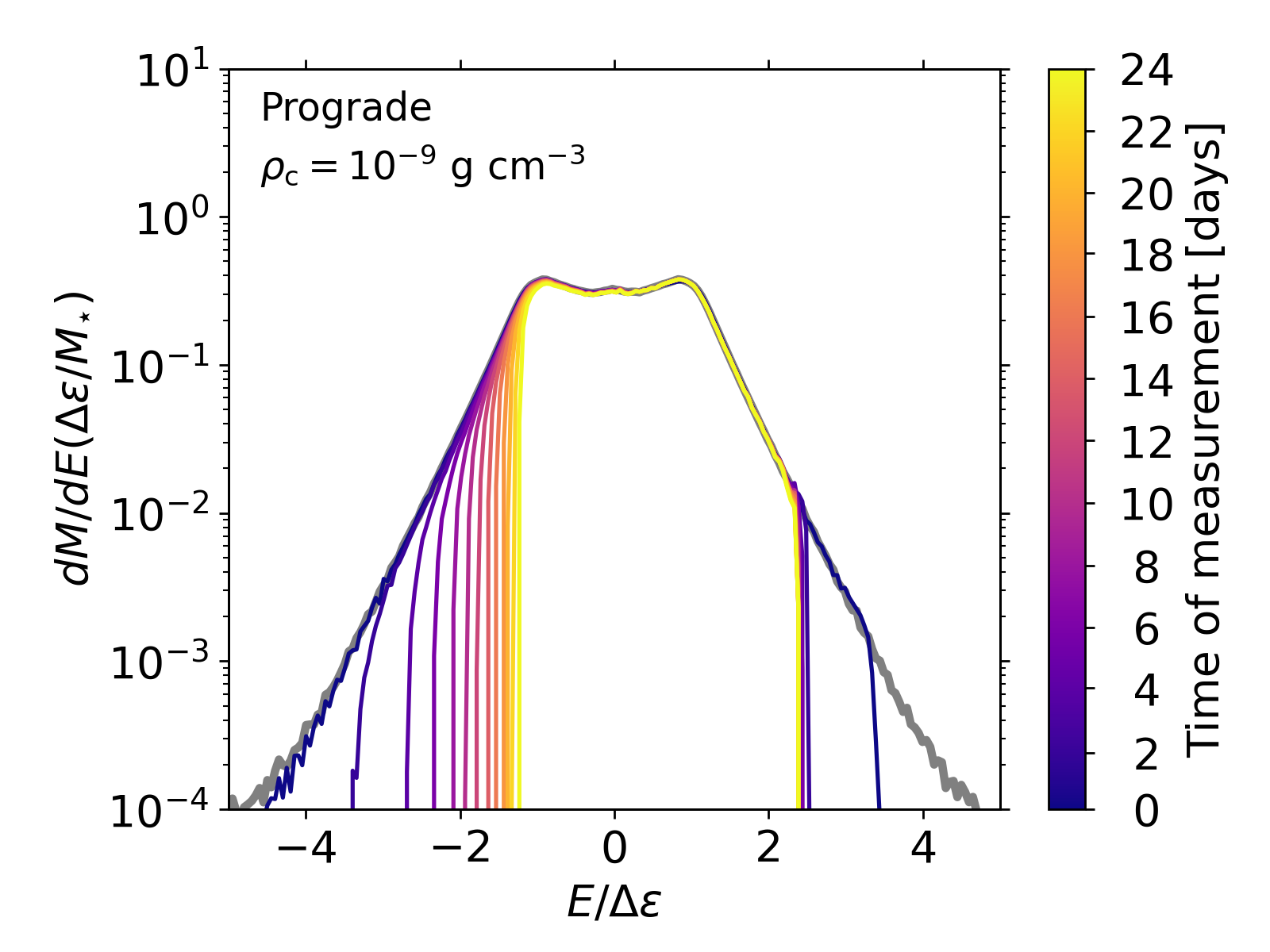}
    \includegraphics[width=8.6cm]{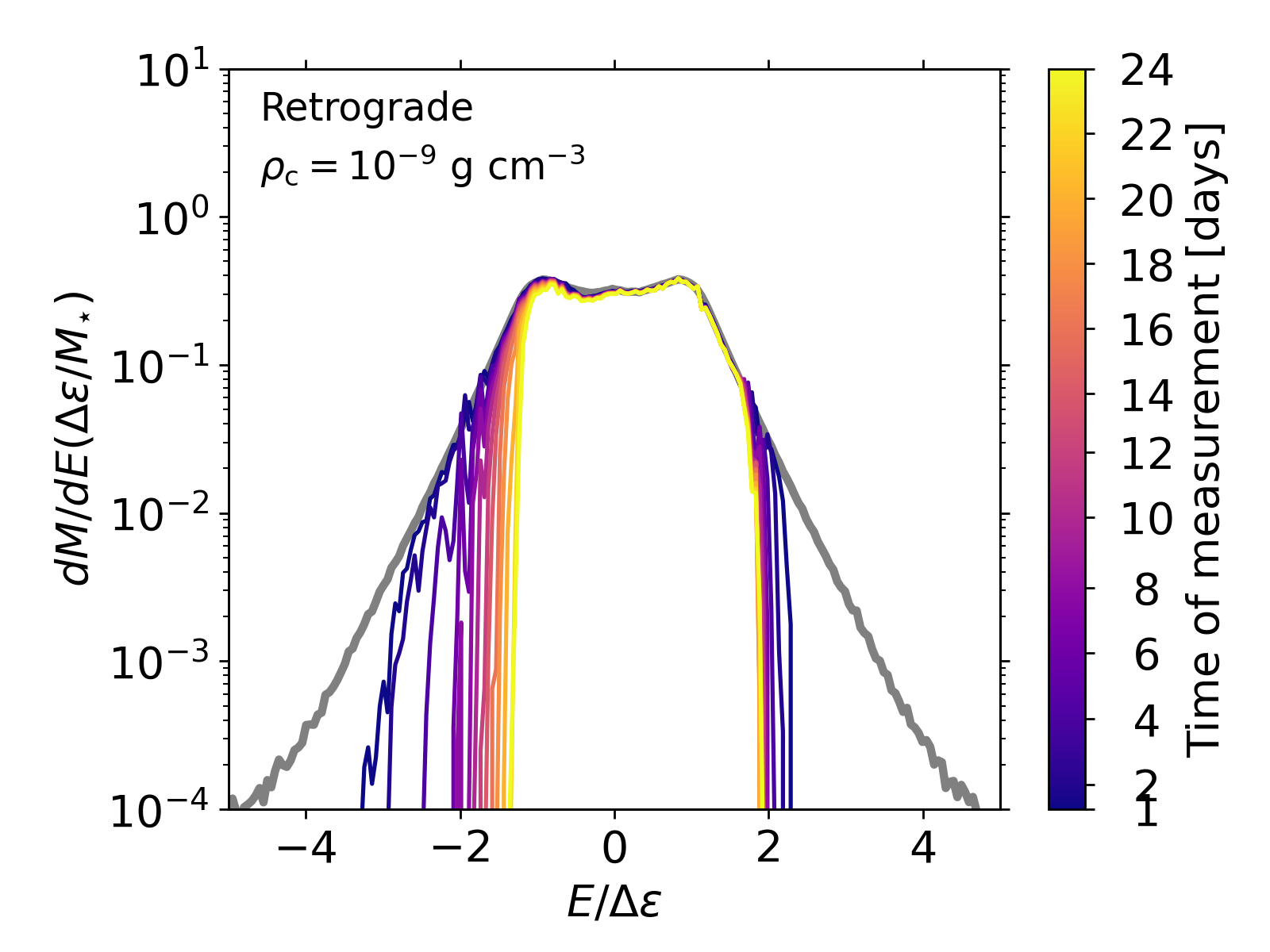}
\caption{Energy distribution of debris produced in a full disruption of a $1\Msol$ star on a prograde (\textit{left}) or a retrograde (\textit{right}) orbit relative to an AGN disk with $\rho_{\rm c}=10^{-7}\gram \cm^{-3}$ (\textit{top}), $10^{-8}\gram \cm^{-3}$, and $10^{-9}\gram \cm^{-3}$, around a $10^{6}\Msol$ SMBH. The grey line in each panel shows TDEs of the same star in vacuum (naked TDEs), which is sitting behind the line for AGN-TDEs at $t\simeq 0$ days. The color bar indicates the time at which the distribution is measured since disruption. Notice different time scales in the color bars. }
	\label{fig:dmde}
\end{figure*}

\begin{figure*}
    \centering
    \includegraphics[width=8.6cm]{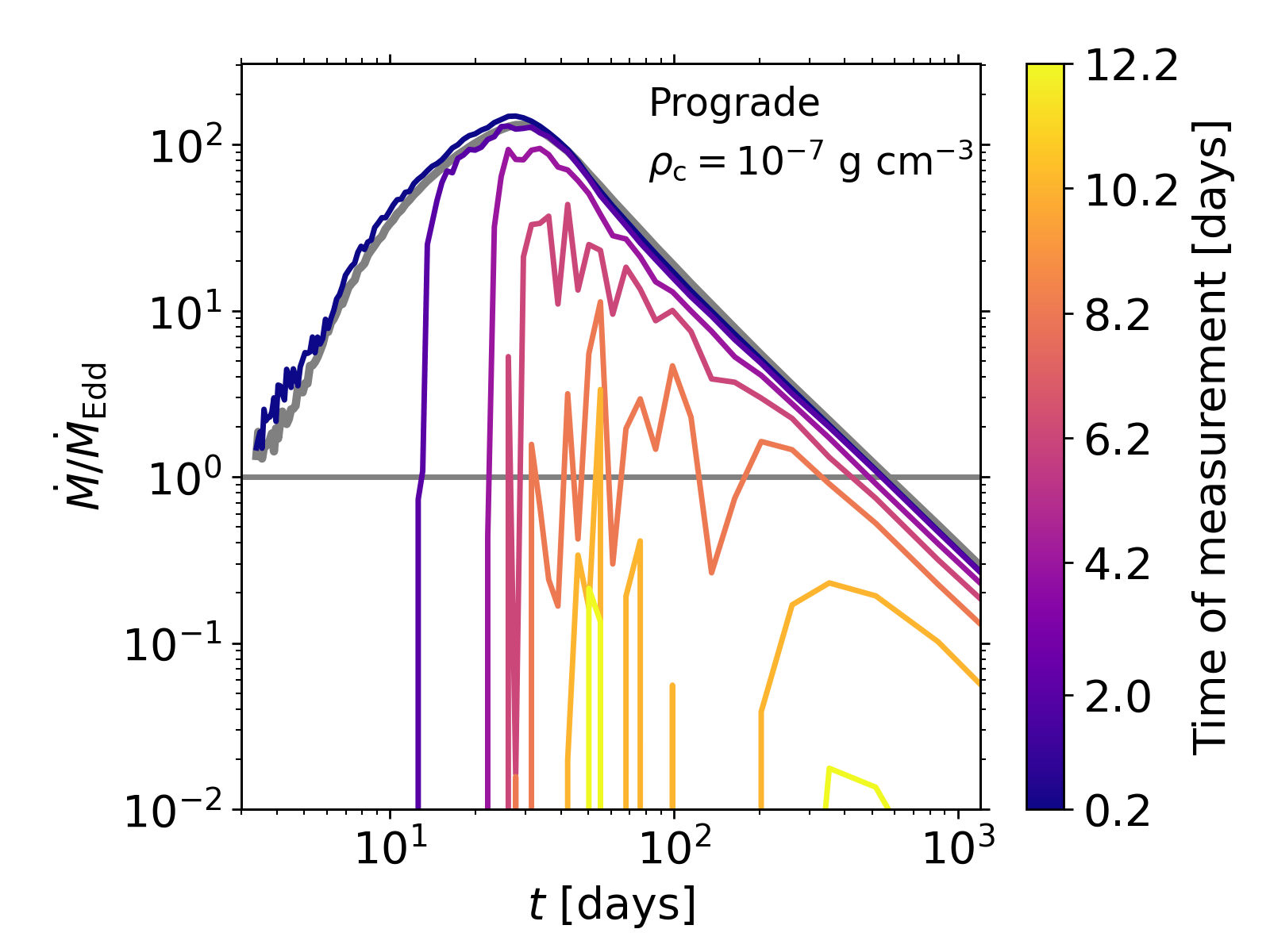}
    \includegraphics[width=8.6cm]{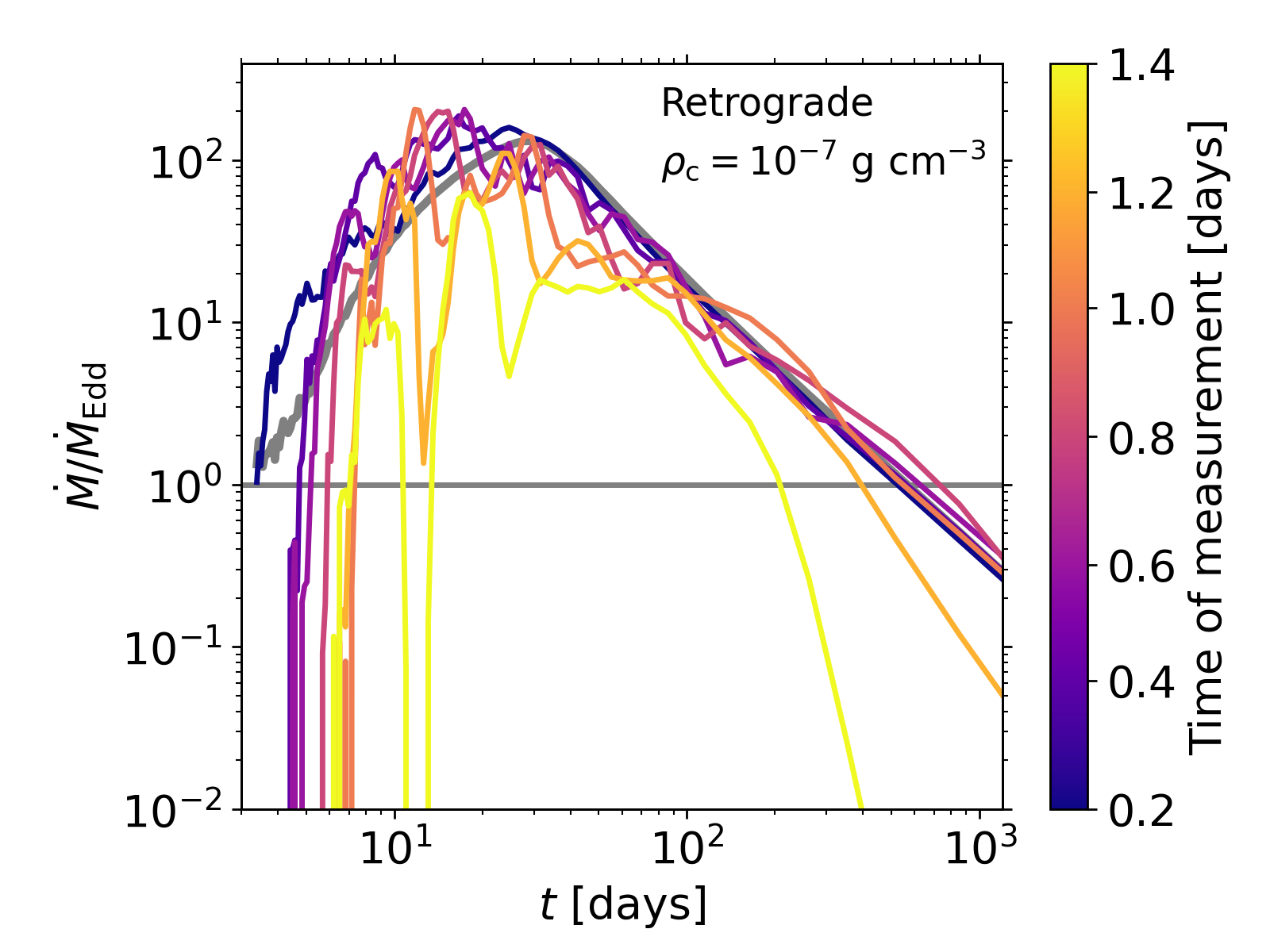}\\
    \includegraphics[width=8.6cm]{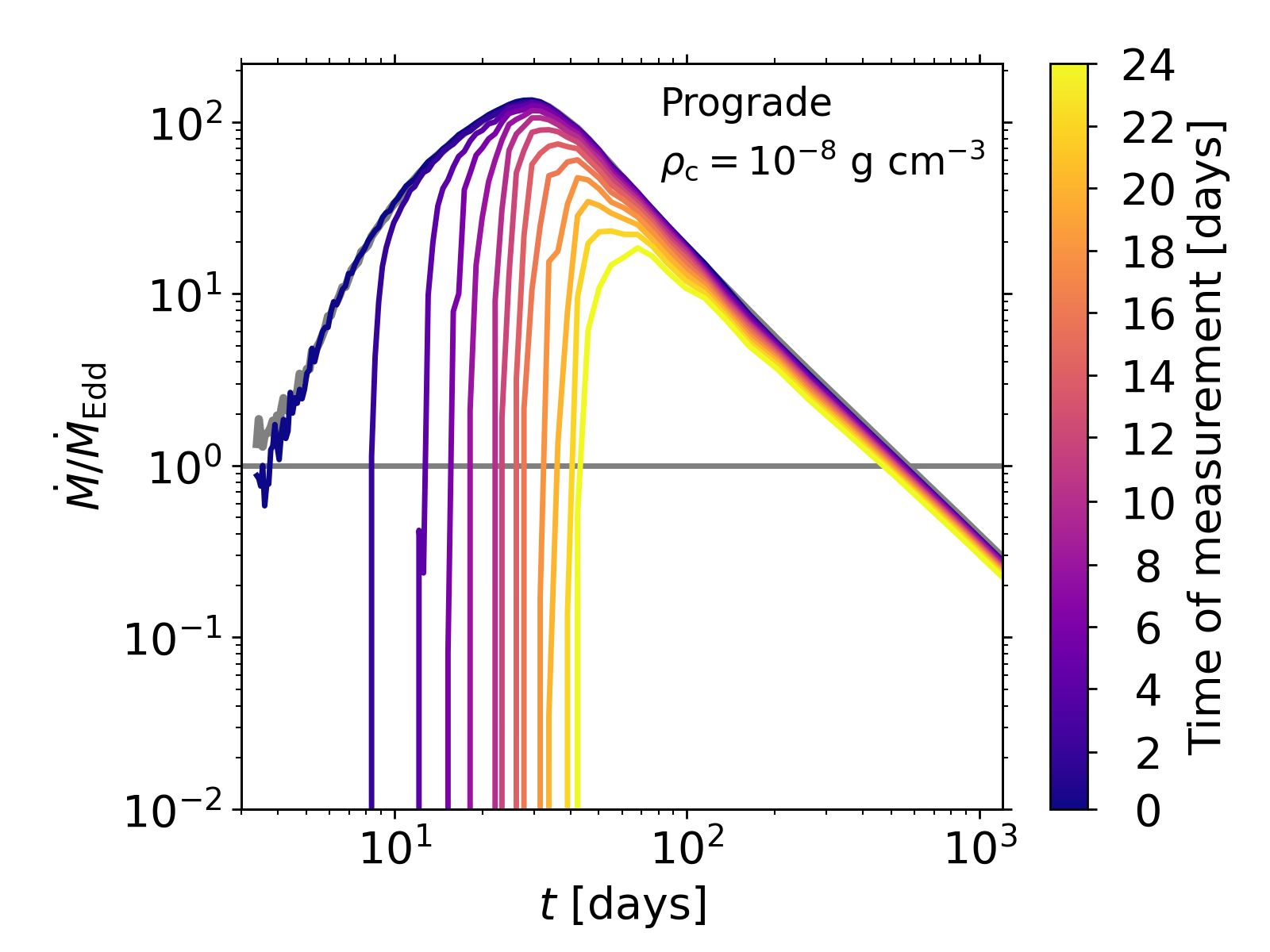}
    \includegraphics[width=8.6cm]{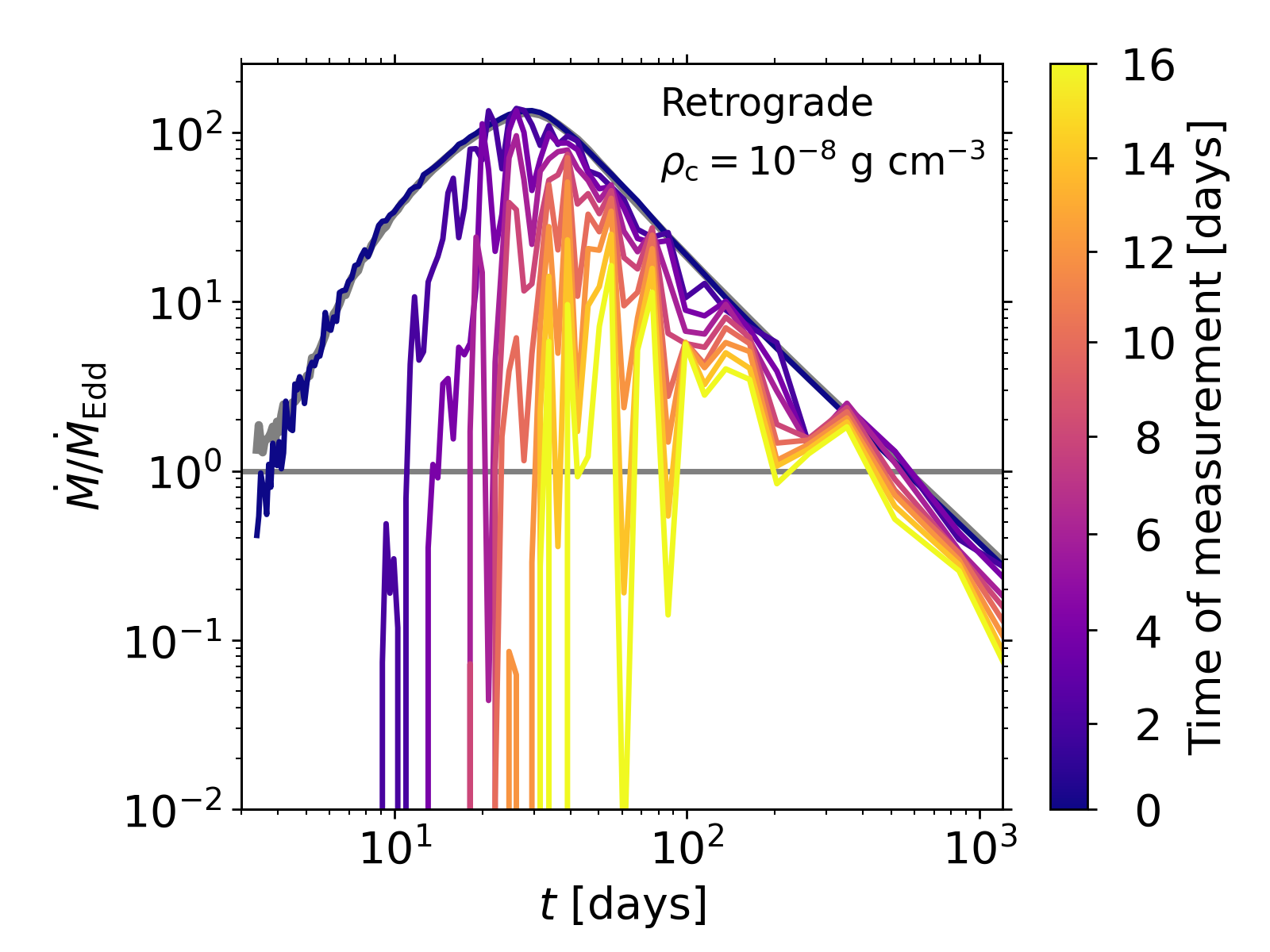}\\
    \includegraphics[width=8.6cm]{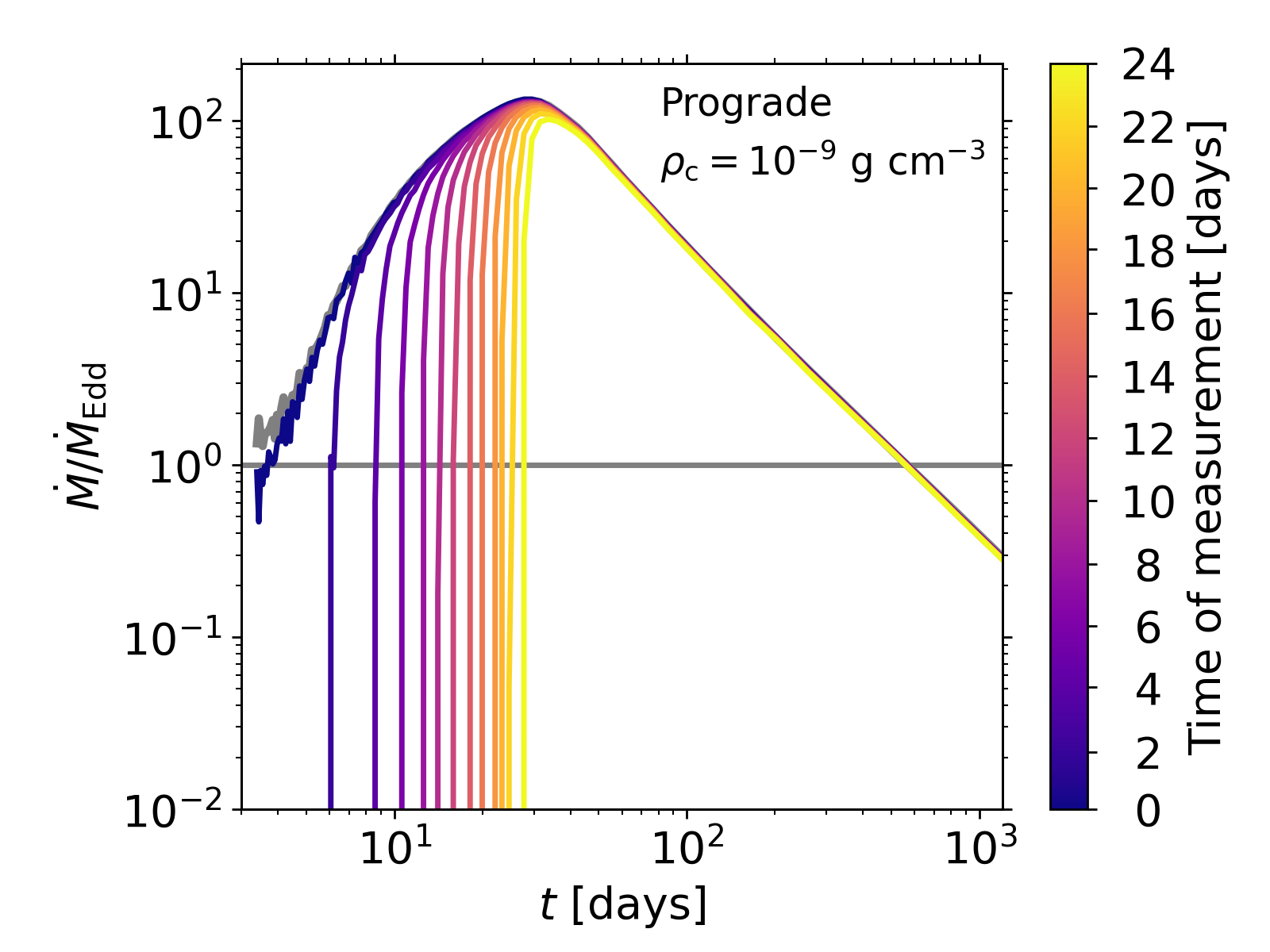}
    \includegraphics[width=8.6cm]{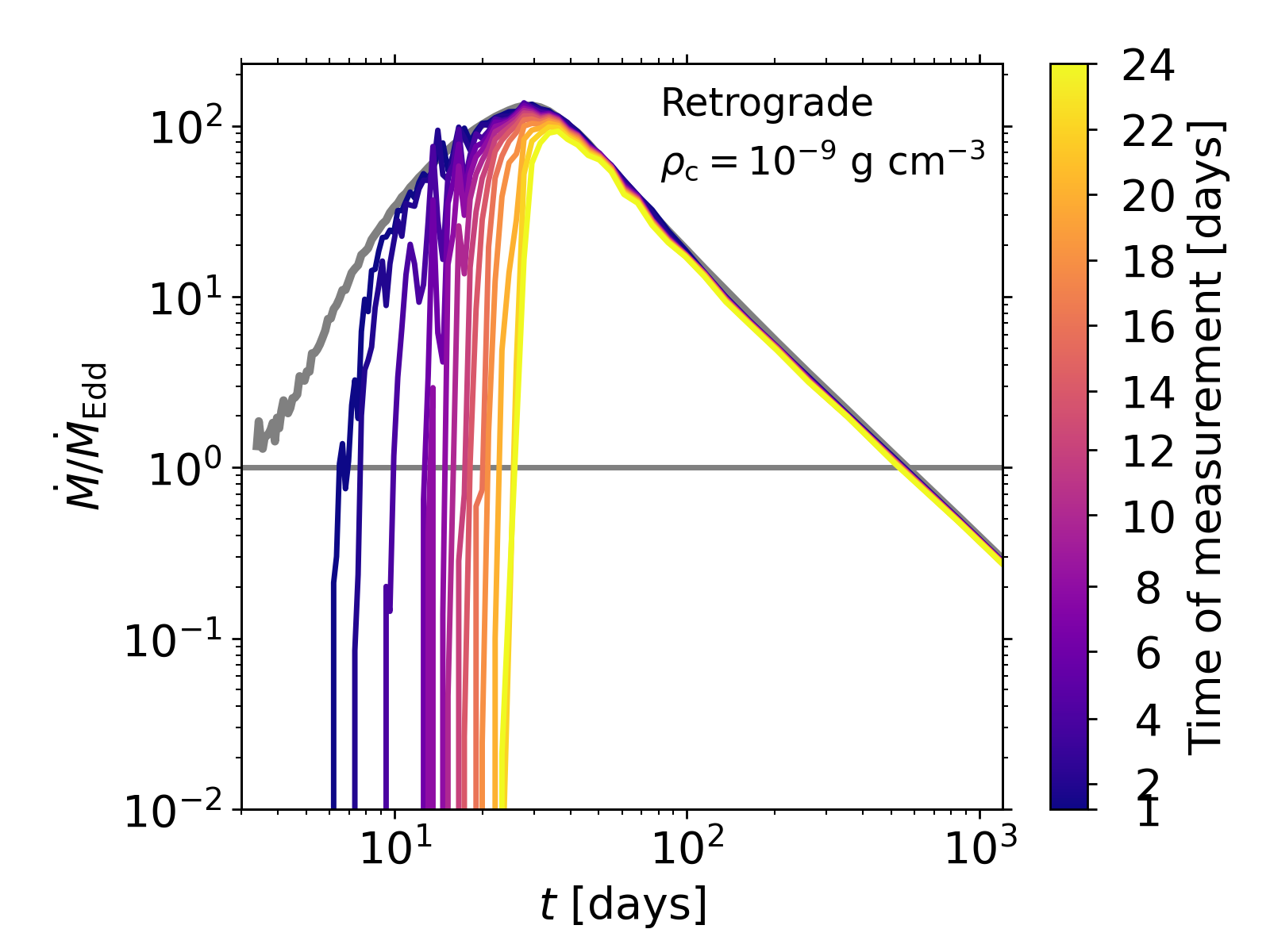}
\caption{Same as Figure~\ref{fig:dmde}, but for the fallback rate. }
	\label{fig:fallback}
\end{figure*}
\subsection{Energy and angular momentum distribution of debris}

In this section, we seek to investigate the energy and energy distribution of debris in AGN disks. Figure~\ref{fig:dlde} presents the distribution $d^{2}M/dEdL$ of specific orbital energy $E$ relative to the SMBH and specific angular momentum $L$ of debris for our fiducial models (i.e., $\mstar=1\Msol$ and $\rho_{\rm c}=10^{-8}\gram\cm^{-3}$). Debris produced in AGN disks undergoes continuous changes in the debris structure, potentially leading to a dramatic modification of the $E-L$ distribution on a relatively short time scale (e.g., before any of the bound material returns to the SMBH). This continuous evolution of the distribution is one of the main differences between TDEs in AGN disks and those in a vacuum (``naked TDEs"). Not long after the disruption (\textit{left} panel, $t=0.2$ days), the $L-E$ distribution appears almost identical between the prograde and retrograde cases. However, as the debris travels away from the SMBH, the distribution continuously evolves differently over time, depending on the relative orientation of the orbit. For the prograde case (\textit{top} panels), one of the most noticeable trends is that the angular momentum increases over time. On the other hand, the angular momentum decreases for the retrograde case (\textit{bottom}  panels). This trend is expected based on how the motion of the debris is aligned or anti-aligned with the disk flow (see Figure~\ref{fig:flowmotion}). Additionally, the distribution for the retrograde case is substantially more perturbed by the disk. At around 10 days, most of the debris in the retrograde case is mixed into the disk, and its angular momentum becomes less than 80\% of the initial angular momentum. Compared to changes in angular momentum within the debris ( $\lesssim 5\%$) for naked TDEs \citep{Cheng2014,Ryu+2020b}, the subsequent change in the angular momentum due to continuous interactions with the disk is much more substantial. Other cases with different disk mid-plane density and stellar masses reveal qualitatively the same trend. 

We further present the distribution of $E$ for $\mstar = 1\Msol$, by integrating $d^{2}M/dEdL$ over $L$, in Figure~\ref{fig:dmde}. Because the energy distribution for the disk with $\rho_{\rm c}\lesssim 10^{-11}$ g cm$^{-3}$ is almost identical to our vacuum case, we only show the distribution for $\rho_{\rm c}\gtrsim 10^{-9}$ g cm$^{-3}$ (also the same for the fallback rate in \S\ref{fig:fallback}).  For comparison, we depict the energy distribution of debris produced in a full disruption in a vacuum sharing the same encounter parameters (grey line in each panel), measured at $2$ days after disruption. For both prograde (\textit{left} panels) and retrograde (\textit{right} panels) cases, the energy distribution at $t=0.2$ days is almost identical to that for the naked TDE, except for the sharp cut-off at the far-end of the wing for the unbound debris, indicating that the most unbound debris has been already mixed to the disk. The subsequent interaction of the debris with the disk gas continuously perturbs the debris starting from its head and tails (where the density is the lowest), corresponding to the wings of the distribution. As a result, the distribution becomes narrower. Notice, however, that the rate at which each side of the distribution becomes narrower is different. At early times ($t\lesssim$ a few days), the unbound debris is lost to the disk at a faster rate than the bound debris. However, the ``mixing" or "slowing-down" rate of the unbound debris becomes slower than that of the bound debris at later times. In all cases except for the retrograde case with $\rho_{\rm c}=10^{-7}\gram \cm^{-3}$, the distribution for the unbound debris does not change at $t\gtrsim 5$ days, while that for the bound debris continues to shrink. This behavior can be understood based on when and how long the debris moves in a denser region of the disk. Upon disruption, the unbound debris advances further out, meaning that it interacts with the disk more at a given time. At later times, once the unbound debris moves beyond the density cusp, because the disk density continues to decrease, the perturbation of the unbound debris due to the disk material becomes increasingly weaker. However, the bound debris stays for a longer time in denser parts of the disk as it slows down before returning to the SMBH, meaning more interactions with the disk.

It is worth noting that the distribution becomes bumpy when irregular debris structure develops due to the Rayleigh–Taylor instability, which is more pronounced for the retrograde case with higher disk densities (see the \textit{right} panel of Figure~\ref{fig:flowmotion}). 

The features mentioned above are also found in the cases with different stellar masses. Given the qualitative similarities, we present the distribution for $\mstar=3\Msol$ and $10\Msol$ in the \textit{upper} panels of Figure~\ref{fig:stellarmass3} and Figure~\ref{fig:stellarmass10}, respectively. Note that at this pericenter distance, the $3\Msol$ star is only partially disrupted and a remnant survives, which correponds to the peak at $E\simeq 0 $ in the energy distribution. 

\subsection{Fallback rate}

Using the energy distribution and assuming a ballistic orbit of the debris, we estimate the mass fallback rate, which is illustrated in Figure~\ref{fig:fallback} for $\mstar=1\Msol$. For completeness, we present the fallback rate for $\mstar=3\Msol$ in  Figure~\ref{fig:stellarmass3} and for $\mstar=10\Msol$ in Figure~\ref{fig:stellarmass10}. The continuous shrinkage of the energy distribution for the bound debris leads to the decrease in the peak fallback rate and increase in the peak fallback time. For $\rho_{\rm c}=10^{-8}\gram\cm^{-3}$ with the prograde orbit (\textit{middle-left} panels), the peak mass return rate decreases from $100\dot{M}_{\rm Edd}$ to $20\dot{M}_{\rm Edd}$ in 24 days. Here, $\dot{M}_{\rm Edd} = L_{\rm Edd}/\eta c^{2}$ where $L_{\rm Edd}$ is the Eddington luminosity with a radiative efficiency $\eta=0.1$. The debris with a bumpy energy distribution in some of the cases (see Figure~\ref{fig:dmde} reveals irregular patterns in the rate on top of the trend of the peak fallback rate and time. For example, for $\rho_{\rm c}=10^{-8}\gram\cm^{-3}$ (\textit{top} panels), the rate curves gradually shift towards the right-bottom corner of the figure while the curves become increasingly bumpy. The bumpiness and the change in the peak fallback rate and time are greater for higher $\rho_{\rm c}$ and for a retrograde orbital configuration. 

One observationally relevant finding is that \textit{the rate at which the bound debris is mixed into the disk is faster than the rate at which the debris returns in all cases shown in Figure~\ref{fig:fallback} ($\rho_{\rm c}\gtrsim 10^{-9}\gram\cm^{-3}$)}. In other words, the bound debris is continuously mixed into the disk before it returns to the SMBH in a coherent fashion like it does in a naked TDE. This suggests that the resulting light curves of AGN TDEs in sufficiently dense gas disks would not simply be TDE-like lightcurves on top of AGN lightcurves (see \S\ref{sub:implication}).

\begin{figure*}
    \centering
    \includegraphics[width=8.6cm]{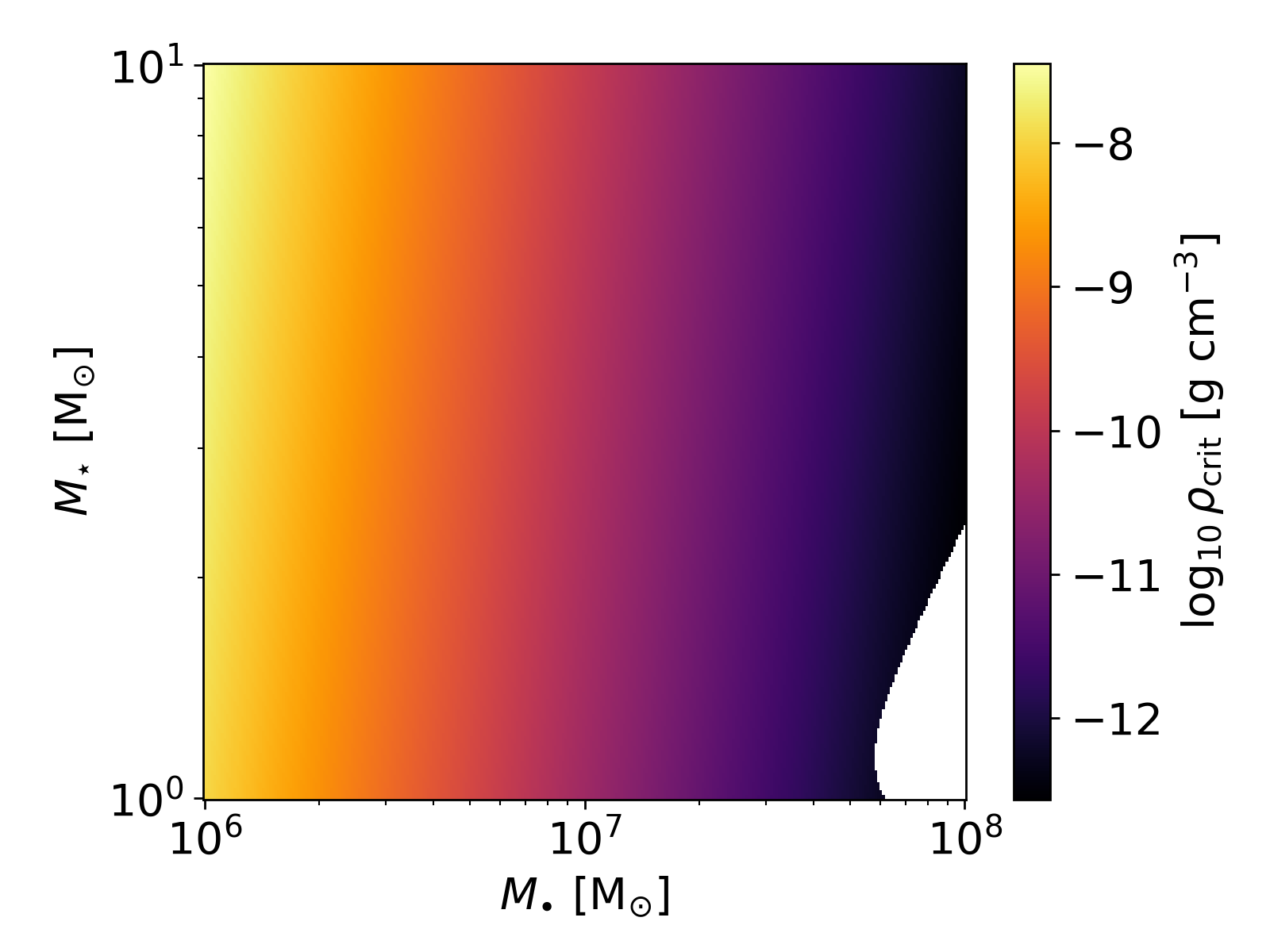}
    \includegraphics[width=8.6cm]{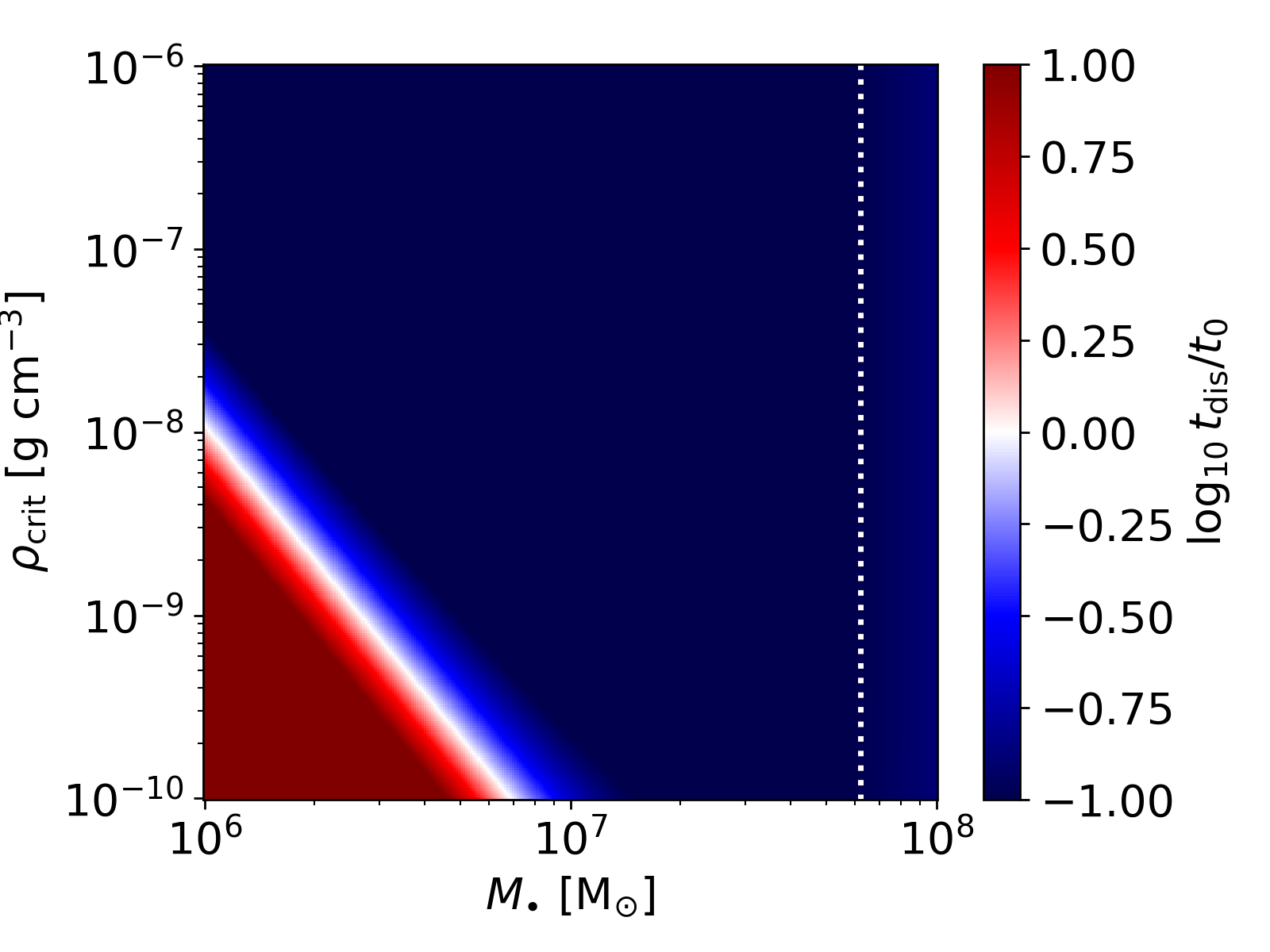}
\caption{The minimum critical density $\rho_{\rm crit}$ (\textit{left}) above which the entire debris is completely mixed into the disk on the peak mass return time $t_{0}$ estimated right after disruption in a $\mstar-\Mbh$ plane and the ratio of the complete dissociation time $t_{\rm dis}$ to the peak mass return time in a $\rho_{\rm crit}-\Mbh$ plane. Here, $t_{0}$ is estimated by including the correction factor for the stellar internal structure and relativistic effects \citep{Ryu+2020a,Ryu+2020b,Ryu+2020c,Ryu+2020d}. The white region near the corner indicates the region of parameter space where stars would be directly captured by the black hole. Similarly, the vertical dashed white line indicates the maximum black hole mass for direct captures.   }
	\label{fig:rhoc_constraint}
\end{figure*}

\section{Discussion}\label{sec:discussion}

\subsection{Lightcurves}\label{sub:implication}

\subsubsection{Passage of star around the SMBH}
A close passage of a star can significantly perturb the inner part of the disk, which possibly enhances the accretion rate until the perturbed disk settles again over a time scale comparable to the cooling timescale. To zeroth order, whether a stellar passage can significantly affect the structure of the inner disk can be measured by comparing the swept-up mass of the star during the pericenter passage, $M_{\rm swept-up}\simeq \rho_{\rm disk}\pi R_{\star}^{2}r_{\rm p}$, to the disk mass within the pericenter distance, $M_{\rm disk}(r<r_{\rm p})\simeq \rho_{\rm disk}(h/r)r_{\rm p}^{3}$, 
\begin{align}
    \frac{M_{\rm swept-up}}{M_{\rm disk}(r<r_{\rm p})} = 0.3 \left(\frac{h/r}{0.05}\right)^{-1}\left(\frac{\mstar}{10\Msol}\right)^{2/3}\left(\frac{\Mbh}{10^{6}\Msol}\right)^{-2/3}\left(\frac{r_{\rm p}/r_{\rm t}}{0.3}\right)^{2}.
\end{align}
This ratio of order unity suggests that a passage of a main-sequence star, in particular a massive one, can significantly affect the inner disk over a very short time scale comparable to the dynamical time at pericenter.
Qualitatively, this may cause a state-change in the inner disk, much like those observed in changing-look AGN exhibiting an increase in luminosity \citep{Graham20}.
The temperature at the innermost stable circular orbit can be parameterized as \citep{McKernan+2022}
\begin{eqnarray}
    T_{\rm ISCO} &\approx& 10^{6}{\rm K} \left(\frac{\Mbh}{10^{6}M_{\odot}}\right)^{-1/4}\left(\frac{\dot{M}}{0.1\dot{M}_{\rm Edd}}\right)^{1/4}\left(\frac{\eta}{0.1}\right)^{-1/4} \\ \nonumber &\times& \left(\frac{r_{\rm ISCO}}{6r_{g}}\right)^{-3/4}\left(\frac{f}{2}\right)^{-1/4},
\end{eqnarray}
where $r_{\rm ISCO}$ is the location of the innermost stable circular orbit and $f$ is a numerical factor. By contrast, the temperature of the shock due to the close pass of the star ($T_{\rm shock} \sim L_{\rm shock}/4\pi R_{\rm shock}^{2} \sigma$) can be parameterized as \citep{McKernan+2022}
\begin{equation}
T_{\rm shock} \approx 4 \times 10^{6}{\rm K} \left(\frac{a}{10r_{g}} \right)^{-3/8} \left(\frac{\rho_{\rm disk}}{10^{-8}{\rm g~cm^{-3}}} \right)^{1/4},
\end{equation}
where we assume $R_{\rm shock} \sim R_{\star}$ and clearly the passage of the star must heat the innermost disk substantially. At such temperatures, prompt X-ray flaring and fast outflows are likely \citep[e.g.][]{Kosec23}. If the heating of passage translates into a fiducial local aspect ratio increase then the puffed-up inner disk is accreted on a shorter viscous timescale $t_{\nu}$. Since $t_{\nu}$ can be parameterized as \citep[e.g.][]{Stern18}
\begin{equation}
t_{\nu} \sim 6~{\rm yr} \left(\frac{h/r}{0.05} \right)^{-2} \left(\frac{\alpha}{0.01} \right)^{-1}\left(\frac{\Mbh}{10^{6}M_{\odot}} \right) \left(\frac{R}{100r_{g}} \right)^{3/2},
\end{equation}
we can see that e.g. a doubling of average disk aspect ratio $h/r$ due to local heating leads to a significantly shorter ($1/4$) accretion timescale and so there is a temporary enhancement in accretion (and therefore $\eta \dot{M} c^{2}$ luminosity) while the local disk accretes and cools over the approximate thermal timescale \citep{Stern18}
\begin{equation}
t_{\rm th} \sim 12 ~{\rm days} \left(\frac{\alpha}{0.01} \right)^{-1}\left( \frac{\Mbh}{10^{6}M_{\odot}}\right) \left(\frac{R}{100r_{g}} \right)^{3/2}.
\end{equation}
Thus, if debris from the TDE can make it back to the SMBH on these timescales ($t_{\rm th}$), the initial impulse heating is continued and added to, in a single episode. If debris from the TDE takes longer than $t_{\rm th}$ to return to the inner disk, then the lightcurve will consist of two separate episodes, the initial perturbation, followed by the debris fallback and accretion. But as explained in the following section, the TDE-like lightcurves from the debris fallback and accretion can only be created when the disk density is sufficiently small. 

\subsubsection{Full tidal disruption event}
The subsequent source of a flare is the stellar debris produced in the tidal disruption. The biggest difference between in-plane AGN TDEs and 'naked' or standard TDEs is the continuous interaction between debris and disk gas, resulting in the time evolution of the debris orbits and debris structure. In naked TDEs, because the debris' orbit is almost ballistic, the post-disruption debris orbit does not change significantly over time until debris returns to the SMBH. This feature allows us to make a prediction for the fallback rate curve \citep{Hills1988,Rees1988} with the energy distribution of debris upon disruption. However, in the case of AGN-TDEs, because the debris continuously interacts with surrounding disk gas, the shape of the fallback rate curve, such as the peak fallback rate and the slope of the decaying part of the fallback rate curve, depends strongly on the disk density ($\rho_{\rm disk}$). If $\rho_{\rm disk}$ is sufficiently high, greater than some critical value ($\rho_{\rm crit}$), the debris is completely mixed with the disk before it ever returns. In this case, no debris returns to the SMBH in a coherent and eccentric fashion as predicted for naked TDEs. 

Using the semi-analytic model developed in \S~\ref{subsec:massloss} we estimate  $\rho_{\rm crit}$ for a complete dissociation of the debris in a retrograde orbit on a time scale comparable to the peak mass return time (estimated assuming naked TDEs) as a function of $\mstar$ and $\Mbh$ in the \textit{left} panel of Figure~\ref{fig:rhoc_constraint}. Although we assume $r_{\rm cusp}=10^{3}r_{\rm g}$ here, $\rho_{\rm crit}$ can be easily calculated using our semi-analytic model (see \S~\ref{subsec:massloss}) with different values of $r_{\rm cusp}$. As shown in the \textit{left} panel, the minimum $\rho_{\rm crit}$ has a relatively weak dependence on $\mstar$ while it mostly depends on $\Mbh$. The reason that $\rho_{\rm crit}$ is lower for higher $\Mbh$ is because each part of the debris travels a longer absolute distance for a given peak mass return time. We also present in the \textit{right} panel the ratio of the time for a complete dissociation $t_{\rm dis}= t (M_{\rm d}=M_{\star})$ of debris to the peak mass return time $t_{0}$ for naked TDEs as a function of $\rho_{\rm crit}$ and $\Mbh$, suggesting that in the large parameter space relevant for AGN disks, debris is completely mixed with the disk before returning to the SMBH.  Thus, for dense AGN disks, \textit{the resulting lightcurves cannot be described by a simple superposition of the luminosity of naked TDEs on top of that of AGN disks}. Conveniently, we can parameterize $\rho_{\rm crit}$ at $t_{\rm dis}\simeq t_{0}$ for the retrograde TDEs, as
\begin{align}
    \rho_{\rm crit} \simeq 10^{-8}{\rm g~cm^{-3}}\left(\frac{\Mbh}{10^{6}\Msol}\right)^{-2.5}.
\end{align}
The value of $\rho_{\rm crit}$ would be higher for prograde TDEs than that for retrograde TDEs. 
Even for those cases where the debris is partially disintegrated, because the mass fallback rate curve would be significantly different from that for naked TDEs, the resulting luminosity associated with the disruption of a star could be different. Nonetheless, the returned debris can perturb the inner disk, which would boost the luminosity temporarily. However, detailed modeling of the response of the disk near the SMBH is beyond the scope of this paper. 

In some cases, disruptions end up adding some mass to the disk without generating a TDE-like flare. However the addition of the mass would have a minimal effect on the disk structure because the disk mass inside the radius at which the debris is completely disintegrated, namely $r\gtrsim10^{3}r_{\rm g}$ for $\rho_{\rm c}\lesssim 10^{-7}{\rm g
~cm}^{-3}$, is much greater than the mass of the debris (see Figure~\ref{fig:debris_mass}). 

Quantitatively, in the limit of low $\rho_{\rm disk} \ll \rho_{\rm c}$ (more specifically $\rho_{\rm disk}\lesssim \rho_{\rm c}/10^{3}$ based on our simulations), AGN-TDEs should look increasingly like standard 'naked' TDEs. Thus, observations of a TDE-like lightcurve in an AGN should indicate a low density disk with $\rho_{\rm disk} \ll \rho_{\rm c}$. A low density disk at large radii might be responsible for late-time radio signatures years post-TDE \citep{Cendes22}, as part of the debris that would otherwise escape interacts with the gas disk and returns later than the main apparently 'naked' TDE. Thus late-time responses to otherwise 'naked' TDEs could indicate either a weak AGN, or a more distant fuel reservoir, interacting with the debris, driving much later material return.

\subsubsection{Partial tidal disruption event}
The source of AGN-TDEs are embedded stars that have either been scattered via dynamical encounters into the AGN loss cone, or on highly eccentric orbits. In both cases, the probability of a partial tidal disruption event (where pericenter passage of the star is close, but not too close, to the SMBH) should be higher than an actual AGN-TDE. It is worth considering the observational implications of partial AGN-TDEs. 

As we have seen above, AGN-TDE debris mixing can be significant, particularly in dense AGN disks. This can inhibit and, if the disk is dense enough ($\rho_{\rm disk}>\rho_{\rm crit}$), completely prevent the return of TDE material to the SMBH. However, in the case of a partial disruption, some of the outer part of the star is stripped, but the core remains coherent. As long as the orbit is bound (e.g. if the orbit is highly eccentric), it should return to the SMBH on approximately the orbital period, $T_{\rm orb} \sim 5 ~{\rm day} (a/600r_{\rm g})^{3/2}(\Mbh/10^{6}\Msol)$ where $a$ is the semimajor axis of the remnant's orbit. Repeated passage of a bound remnant will generate similar heating of the inner disk to the first pericenter passage. Such partial disruption perturbations could yield transients like quasi-periodic eruptions (QPEs) as observed in X-rays in some AGNs around low mass SMBHs \citep[e.g.][]{Wevers22}. The magnitude of any repeating flare depends on the remnant mass and the ratio of the local thermal timescale ($t_{\rm th}$) to the returning timescale ($T_{\rm orb}$ in this case). For example, if the remnant mass is smaller and $t_{\rm th}>T_{\rm orb}$, the heating flare will appear less prominent against a higher AGN continuum state. In order to explain observed QPE timescales of O(day), this would require highly eccentric retrograde stellar orbits at $a \sim {\rm few} \times 10^{2}r_{g}$, around smaller mass SMBHs. Such orbits may occur early on in the AGN phase due to disk capture \citep{Yihan23} or rapid retrograde orbital decay \citep{McKernan+2022}.

\subsection{Metallicity of AGN disks}\label{sub:implication}
AGNs are generally believed to have high  metallicity. In particular the broad line region (BLR) metallicity in AGNs is observed to be substantially super-solar out to high redshift  \citep{HamannFerland99,Juarez09}. Quasar host galaxies at $z<2$ are surrounded by metal-enriched cool gas, believed to originate in AGN outflows \citep{Prochaska13}. Of course some of this metallicity enrichment could come from supernovae embedded in AGN at a rate of $O(10^{-4}){\rm yr^{-1}}$ \citep{Juarez09}, a rate which is very similar to the expected standard TDE rate. It is unclear how stars embedded in AGN disks evolve; but it is possible that they do not undergo supernovae but instead grow in mass and support themselves by inflow of fresh hydrogen from the AGN disk \citep{Cantiello+2021,jermyn22}. If this occurs non-negligibly often, then supernovae would be more rare in AGNs than naively expected from standard stellar evolution, making TDEs a plausible means of enriching AGN metallicity. TDEs can also occur around stellar mass BH embedded in AGN disks, yielding micro-TDEs \citep{Hagai16,Yang22}. Such micro-TDEs can also contribute to metallicity enhancement in the disk. 

Assuming TDEs are the sole source of metallicity enhancement through the mixing of stellar debris with the disk, one can estimate, to an order of magnitude, the number of AGN-TDEs, denoted by $N_{\rm TDE}$, of stars with metallicity $Z_{\star}>Z_{1}$ required to elevate the metallicity of the AGN disk from $Z_{0}$ to $Z_{1}$. The total enclosed mass of the disk within a 
 distance at which a fractional mass $\xi$ of the debris is mixed after $N_{\rm TDE}$ TDEs is $\xi N_{\rm TDE}\mstar + M_{\rm disk}$ and the total mass of metals after TDEs $Z_{0}M_{\rm disk} + Z_{\star}\xi N_{\rm TDE}\mstar$. Assuming the total enclosed mass of the disk is conserved, $N_{\rm TDE}$ can be expressed as, 
\begin{align}
    N_{\rm TDE} &= \left(\frac{Z_{1}-Z_{0}}{Z_{\star}-Z_{1}}\right) \left(\frac{M_{\rm disk}}{\xi\mstar}\right),
\end{align}
where $M_{\rm disk}$ is the enclosed mass of the disk into which debris with a mass of $\xi\mstar$ is mixed. As an example, to enhance the metallicity from $Z_{2}\simeq 0.1Z_{\odot}$ to $Z_{1}\simeq 0.9Z_{\odot}$ through TDEs of $1\Msol$ stars, $N_{\rm TDE}\simeq 1000\xi^{-1}(M_{\rm disk}/10^{2}\Msol)(M_{\star}/\Msol)^{-1}$ when $Z_{\star}=Z_{\odot}$. If $Z_{\star}=2Z_{\odot}$, $\simeq 100\xi^{-1}(M_{\rm disk}/10^{2}\Msol)(M_{\star}/\Msol)^{-1}$. 

However, it is important to note that many variables, namely $\mstar$, $Z_{\star}$, $M_{\rm disk}$, and $\xi$, are highly uncertain. To determine these quantities accurately, a more detailed modeling of the dynamics in stellar clusters around AGN disks would be necessary.

\section{Caveats}\label{sec:caveats}

Although our simulations treat hydrodynamical effects accurately, there are two main caveats. First, no relativistic effects are included. It has been recognized that relativistic effects would play a major role in determining the evolution of debris in TDEs by massive black holes \citep[e.g.,][]{BonnerotStone2021}. This would be applicable to in-plane TDEs in AGN disks with low disk densities. However, if the disk density is sufficiently high so that the debris is mixed before its return, relativistic effects on the long-term evolution of debris would be irrelevant. However, because debris would stay a longer time near pericenter, it is possible that the perturbation of the inner disk at the first pericenter passage would be stronger for more relativistic cases (e.g., extremely relativistic TDEs, \citealt{Ryu+2023}). Second, we do not include radiation pressure in our simulations. The standard AGN disk model suggests that the inner part of the disk is radiation pressure-dominated \citep{ShakuraSunyaev1973,SirkoGoodman2003}. Because our disk is supported only by the gas pressure, the temperature is significantly higher than the case where the disk is supported both by the radiation and gas pressure. We will investigate the impact of radiation pressure on the disk temperature profile and how this in turn affects the time evolution of the debris in a follow-up project.

\section{Summary and Conclusions}\label{sec:conclusion}

In this work, we investigated the evolution of debris in tidal disruption events of main-sequence stars by a $10^{6}\Msol$ supermassive black hole, surrounded by a gaseous disk, using the moving-mesh hydrodynamics simulation code {\small AREPO}. We consider three stellar masses, $\mstar=1\Msol$, $3\Msol$, and $10\Msol$, and a range of mid-plane maximum disk densities $\rho_{\rm c}= 10^{-12}-10^{-7}{\rm g~ cm}^{-3}$. 

The results of the simulations can be summarized as follows,\\
\begin{enumerate}
    \item Stellar debris produced in an in-plane disruption in an AGN disk is continuously perturbed by the disk gas as it plows through the disk. As a result, the energy and angular momentum of the debris evolves differently over time relative to TDEs in a vacuum. For prograde TDEs (those of stars on a prograde orbit relative to the disk orbit), the debris' angular momentum increases whereas the debris' angular momentum decreases for retrograde TDEs. 

    \item For a sufficiently high disk density, a large fraction of the debris can be disintegrated and mixed into the disk via the interaction with the disk. The mixing is more significant for retrograde TDEs. This gradual mixing is a unique feature that is clearly distinguished from tidal disruption events in a vacuum. 

    \item For high density AGN disks it is likely that the debris produced in retrograde TDEs is fully mixed into the disk before any of the debris material returns in a coherent fashion as in naked TDEs. The critical density above which the retrograde debris is completely mixed into the disk on a timescale comparable to the peak mass return time for naked TDEs is $\rho_{\rm crit} \sim 10^{-8}{\rm g~cm}^{-3}(\Mbh/10^{6}\Msol)^{-2.5}$. Note that this critical density is the minimum density of disintegration of the entire debris. The density above which no bound material returns to the SMBH would be lower, $\rho_{\rm crit, bound} \sim 10^{-9}{\rm g~cm}^{-3}(\Mbh/10^{6}\Msol)^{-2.5}$. Even for prograde TDEs, no coherent fallback has been found when the maximum disk density is $\geq 10^{-9}{\rm g~cm}^{-3}$.

    \item The mixing of the stellar material into the disk has several astrophysical implications. First, the light curves of in-plane TDEs in an AGN disk, whose density is high enough to cause the disintegration of the debris, could be significantly different from just the superposition of light curves of naked TDEs on top of those of AGN disks. A first burst should originate from the close passage of a star, perturbing and heating up the inner part of the disk, possibly resulting in an enhancement of the accretion rate until the perturbed disk settles on the local thermal timescale ($t_{\rm th}$). Thus we expect an observable state-change in the AGN emission from a close passage, similar to that observed in changing-look AGN, with significant X-ray flaring around smaller mass SMBHs. At low-density ($\rho_{\rm disk} \lesssim \rho_{\rm crit, bound}/10^{2}$), the gas reservoir at moderately large distances from the SMBH ($>10^{4}r_{\rm g}$) might generate late-time debris return and account for recent \textit{very} late-time signatures of (otherwise naked) TDEs \citep[e.g.][]{Cendes22}. At modest disk densities ($ \rho_{\rm crit, bound}/10^{2}<\rho < \rho_{\rm crit, bound}$), AGN TDE debris is only partially mixed and the rest returns, generating a secondary, longer, inner disk flaring episode via both inner disk perturbation as well as accretion of debris. At high disk densities ($\rho_{\rm disk} \geq \rho_{\rm crit,bound}$) no TDE-like light curves are created, past the initial state-change, but a mildly elevated accretion rate (and higher luminosity) should persist for years due to the complete mixing of debris.  
    
    \item For partial disruptions, recurring  passage of the stellar remnant will heat the inner disk, creating recurring flares. A population of highly eccentric retrograde orbiters around low mass SMBHs should produce quasi-periodic eruptions (QPEs) in X-rays from partial AGN TDEs. Whether each flare is separate or the disk response is blended depends on the orbital period of the embedded partially disrupted star as well as disk properties. Short-timescale QPEs are therefor a test of both the  population of eccentric retrograde orbiters and the AGN disk.

    \item The mixing of the stellar debris in AGN TDEs contributes to the enhancement of the disk metallicity. Supernovae of stars embedded in a disk could be another source of metallicity enhancement. But if stars grow in mass without undergoing supernovae, mixing from AGN TDEs (and micro-TDEs) could be an important mechanism to elevate disk metallicity. 

\end{enumerate}
    
    Identifying such events in a large sample of AGN can provide a constraint on typical densities of AGN disks and the embedded stellar population while the disk exists which are otherwise are hard to probe. Going forward, it will be imperative to understand the shape of the lightcurves that account for the passage of the star and the evolution of debris as a function of disk density. It will also be important to track returning masses in the case of partial AGN TDEs, to test models of QPEs in AGN around low mass SMBHs. In order to investigate the contribution of AGN TDEs to the evolution of metallicity enhancements in AGN disks, detailed modeling of the dynamics between stellar-mass objects in nuclear star clusters surrounding AGN disks would be required over a cosmological time scale. 

\section*{Acknowledgements}

TR is very grateful to Max Gronke for a useful discussion for the mixing of gas. This research project was conducted using computational resources (and/or scientific computing services) at the Max-Planck Computing \& Data Facility. The authors gratefully acknowledge the scientific support and HPC resources provided by the Erlangen National High Performance Computing Center (NHR@FAU) of the Friedrich-Alexander-Universität Erlangen-Nürnberg (FAU) under the NHR project b166ea10. NHR funding is provided by federal and Bavarian state authorities. NHR@FAU hardware is partially funded by the German Research Foundation (DFG) – 440719683. In addition, some of the simulations were performed on the national supercomputer Hawk at the High Performance Computing Center Stuttgart (HLRS) under the grant number 44232. BM \& KESF are supported by NSF AST-2206096 and NSF AST-1831415 and Simons Foundation Grant 533845, with additional sabbatical support from the Simons Foundation.  NWCL gratefully acknowledges the generous support of a Fondecyt General grant 1230082, as well as support from Millenium Nucleus NCN19\_058 (TITANs) and funding via the BASAL Centro de Excelencia en Astrofisica y Tecnologias Afines (CATA) grant PFB-06/2007.  NWCL also thanks support from ANID BASAL project ACE210002 and ANID BASAL projects ACE210002 and FB210003. 

\section*{Data Availability}
Any data used in this analysis are available on reasonable request from the first author.

\bibliographystyle{mnras}

\appendix

\section{Energy distribution and fallback rate for different stellar masses}\label{appendix1}

\begin{figure*}
    \centering
    \includegraphics[width=7.cm]{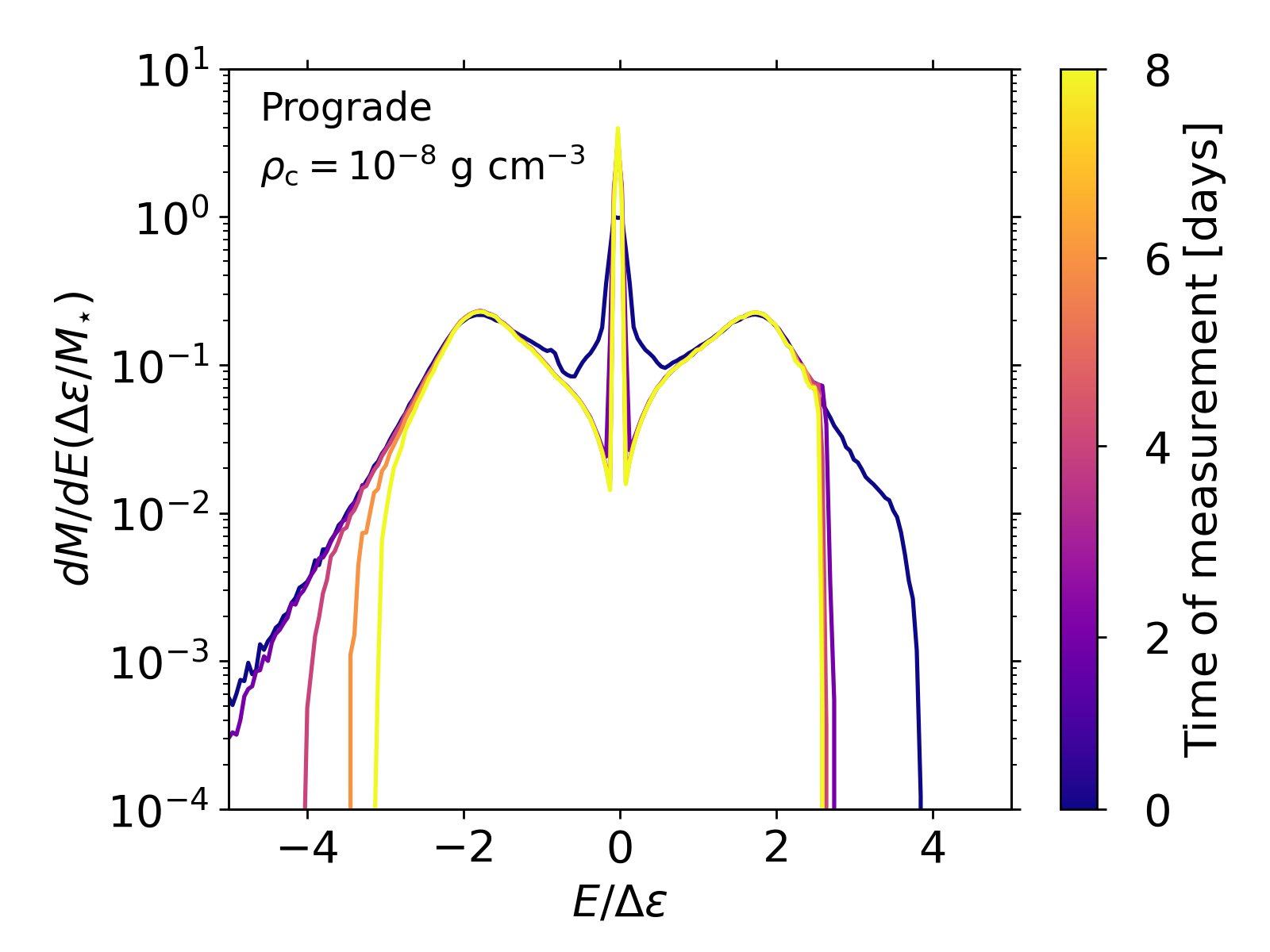}
    \includegraphics[width=7.cm]{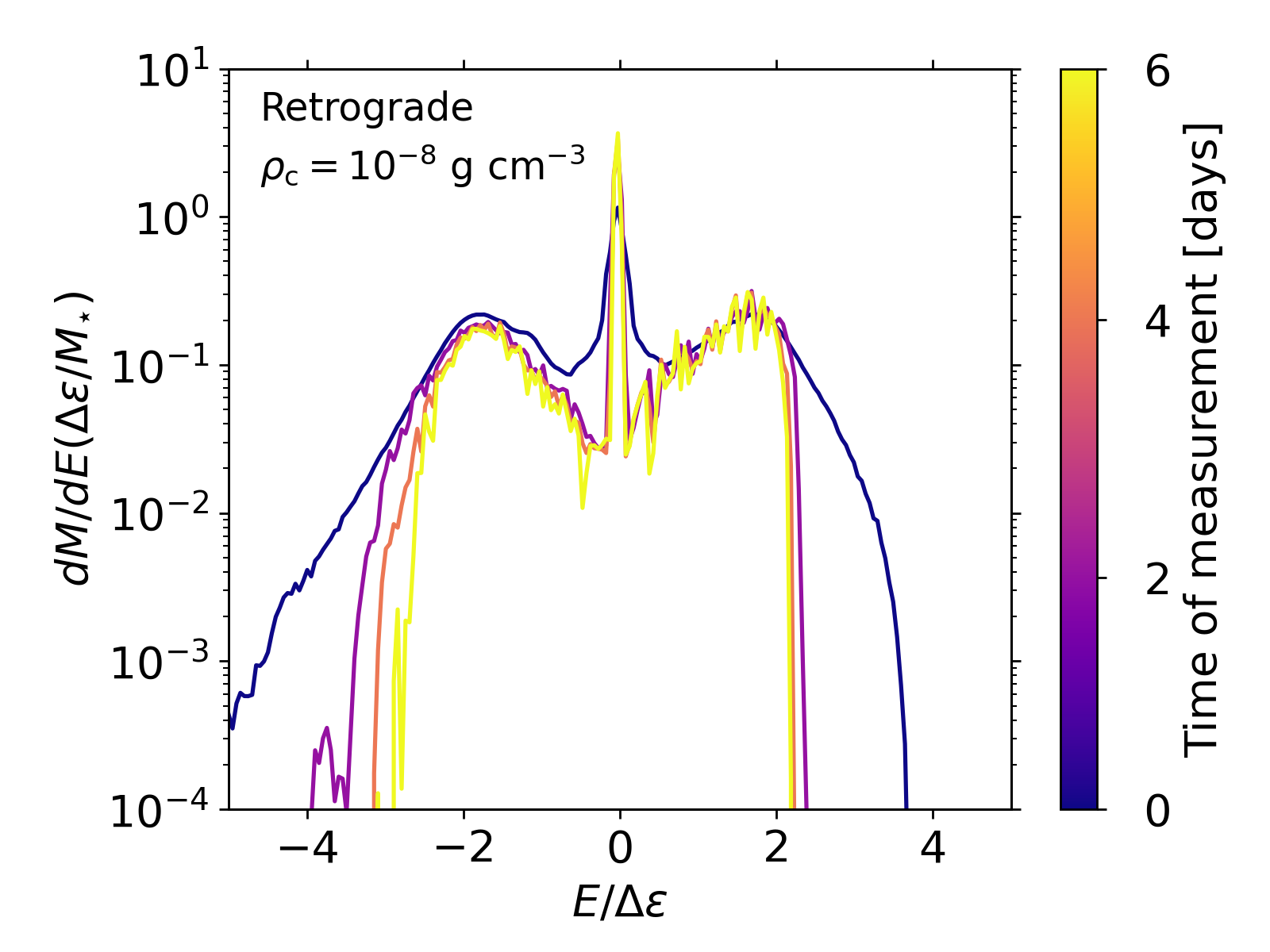}\\
    \includegraphics[width=7.cm]{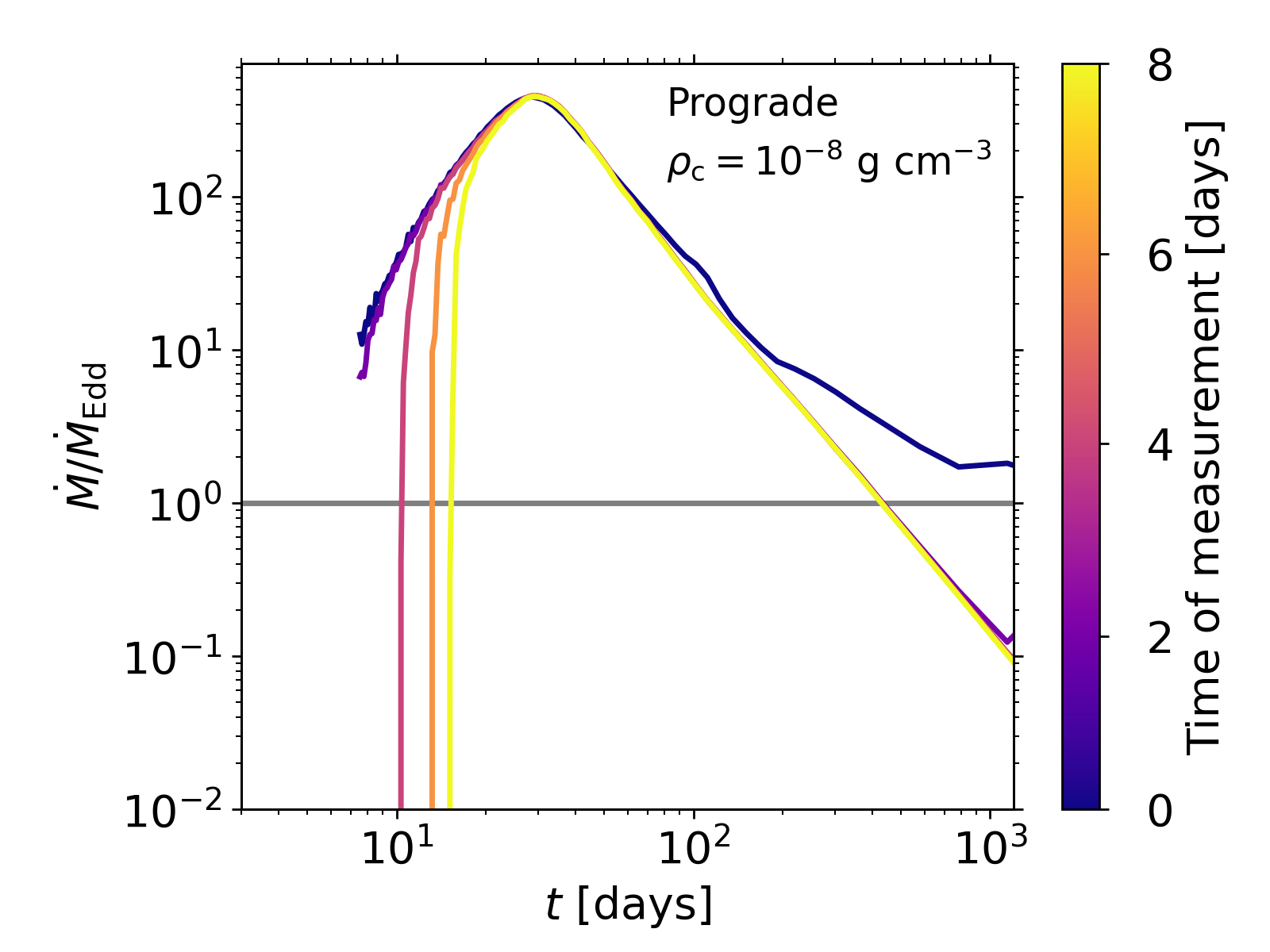}
    \includegraphics[width=7.cm]{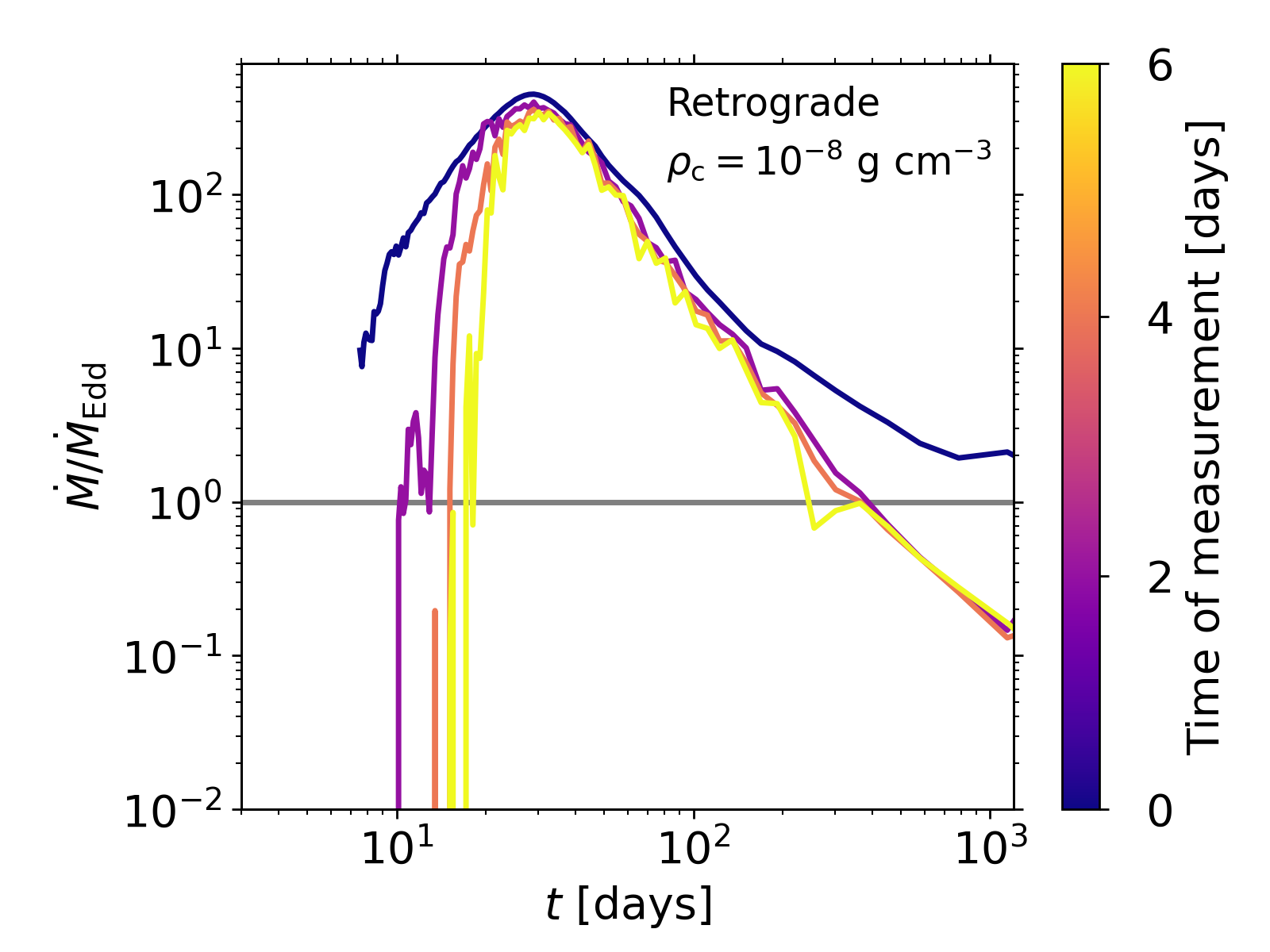}
\caption{Energy distribution (\textit{top}) and fallback rate (\textit{top}) for $\mstar=3\Msol$ on a prograde (\textit{left}) and retrograde (\textit{right}) orbit. Note that the peak of the distribution at $E\simeq 0$ indicates that a remnant survives after disruption. }
	\label{fig:stellarmass3}
\end{figure*}

\begin{figure*}
    \centering
    \includegraphics[width=7.cm]{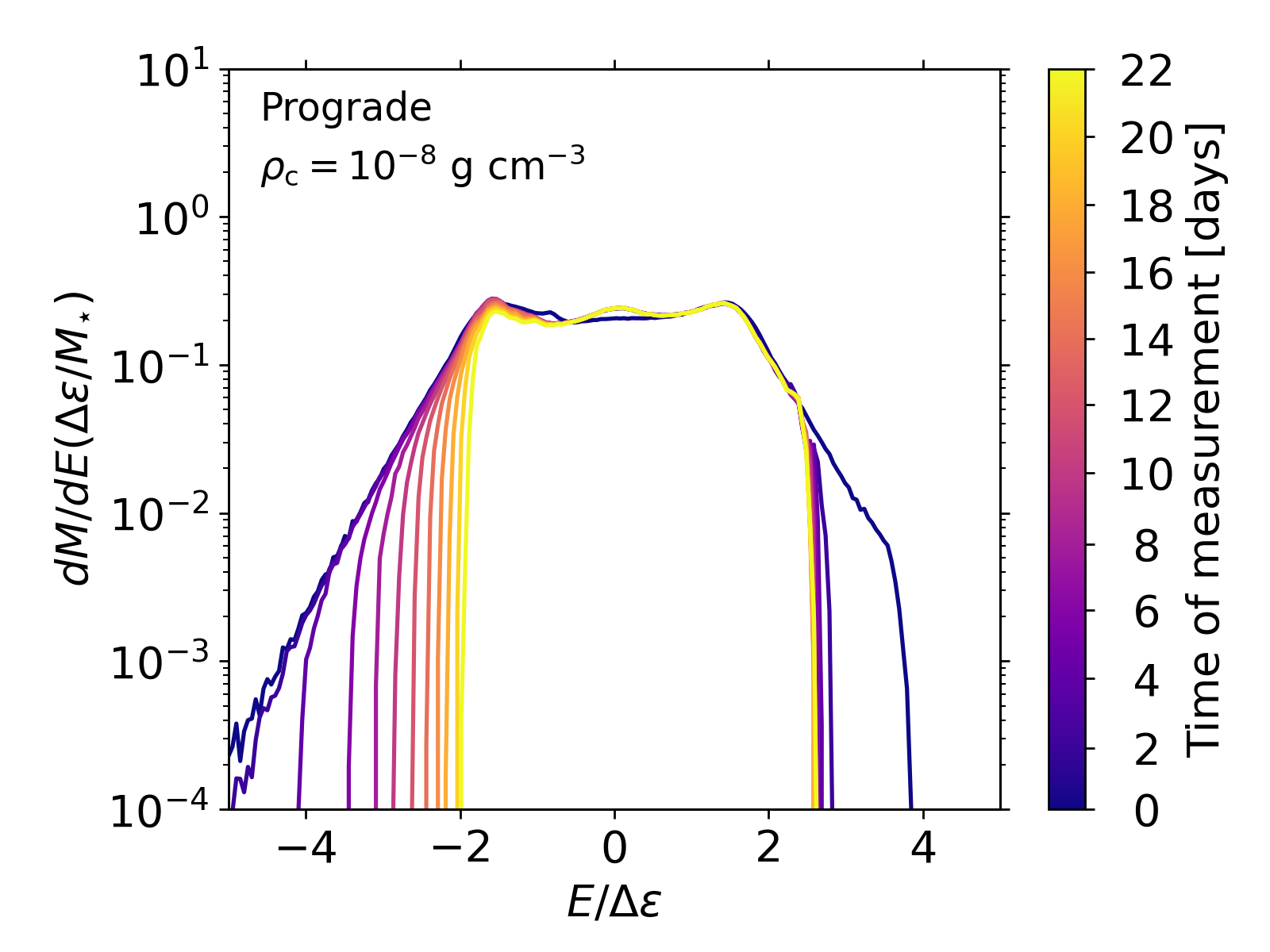}
    \includegraphics[width=7.cm]{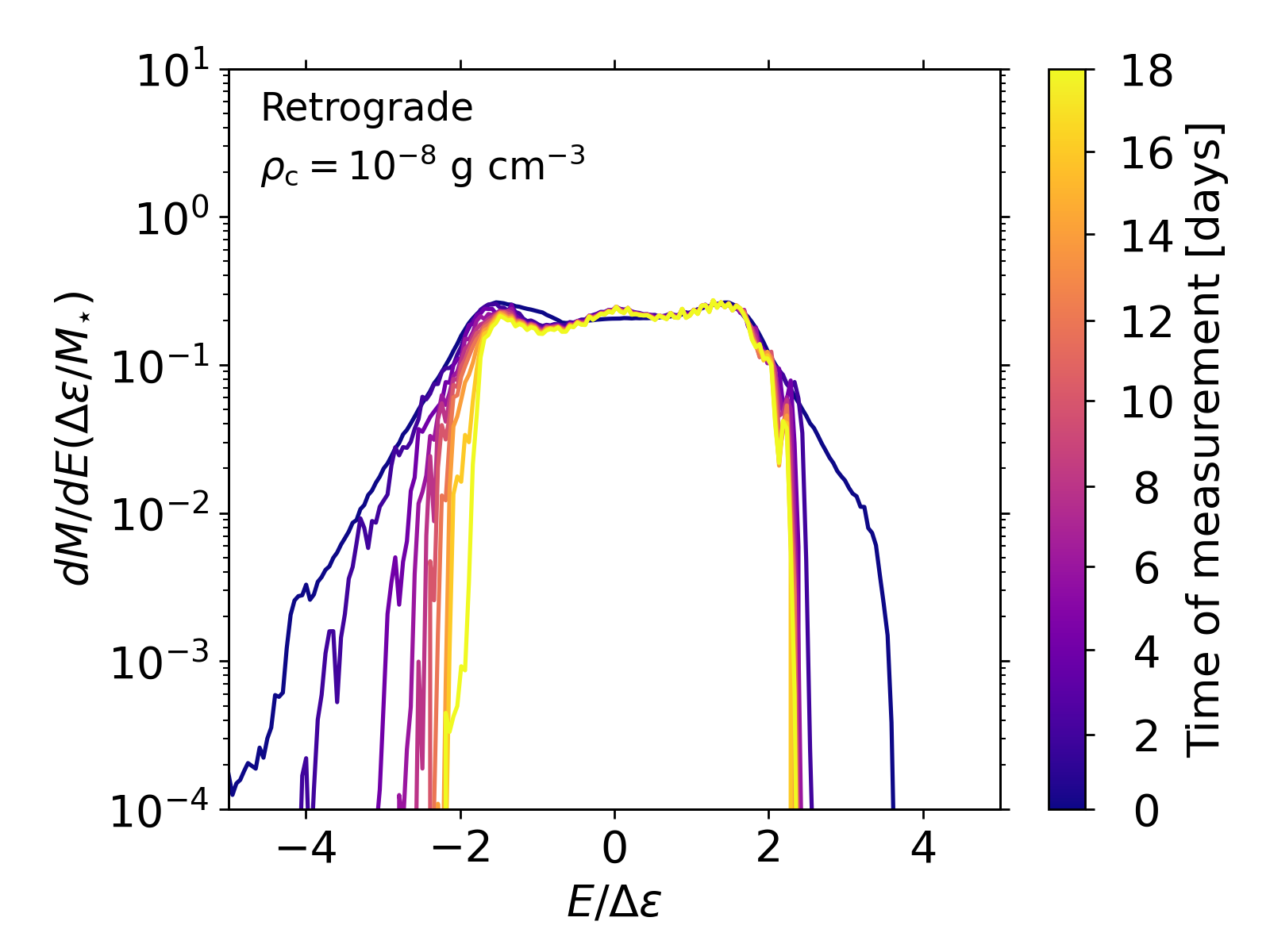}\\
    \includegraphics[width=7.cm]{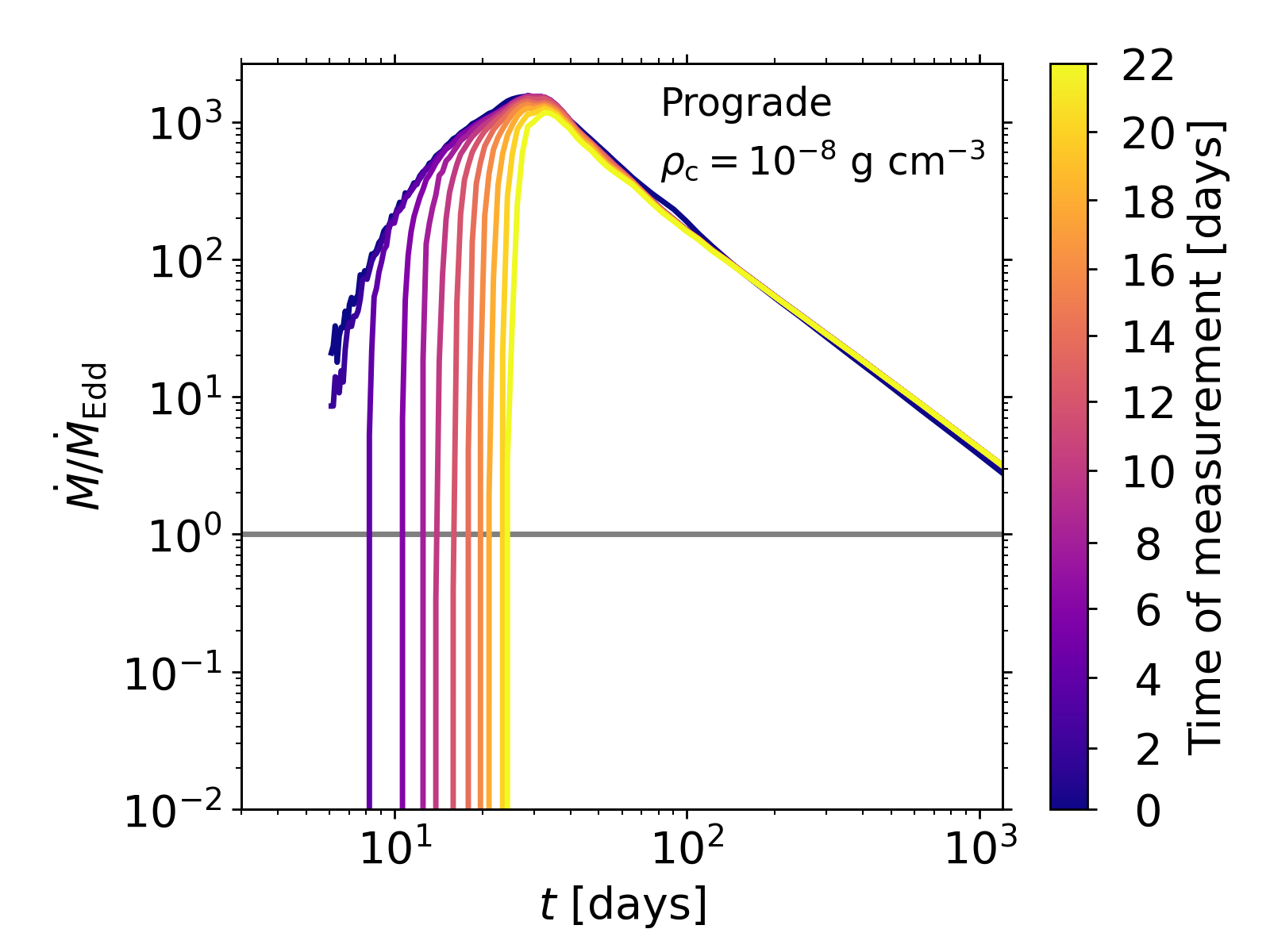}
    \includegraphics[width=7.cm]{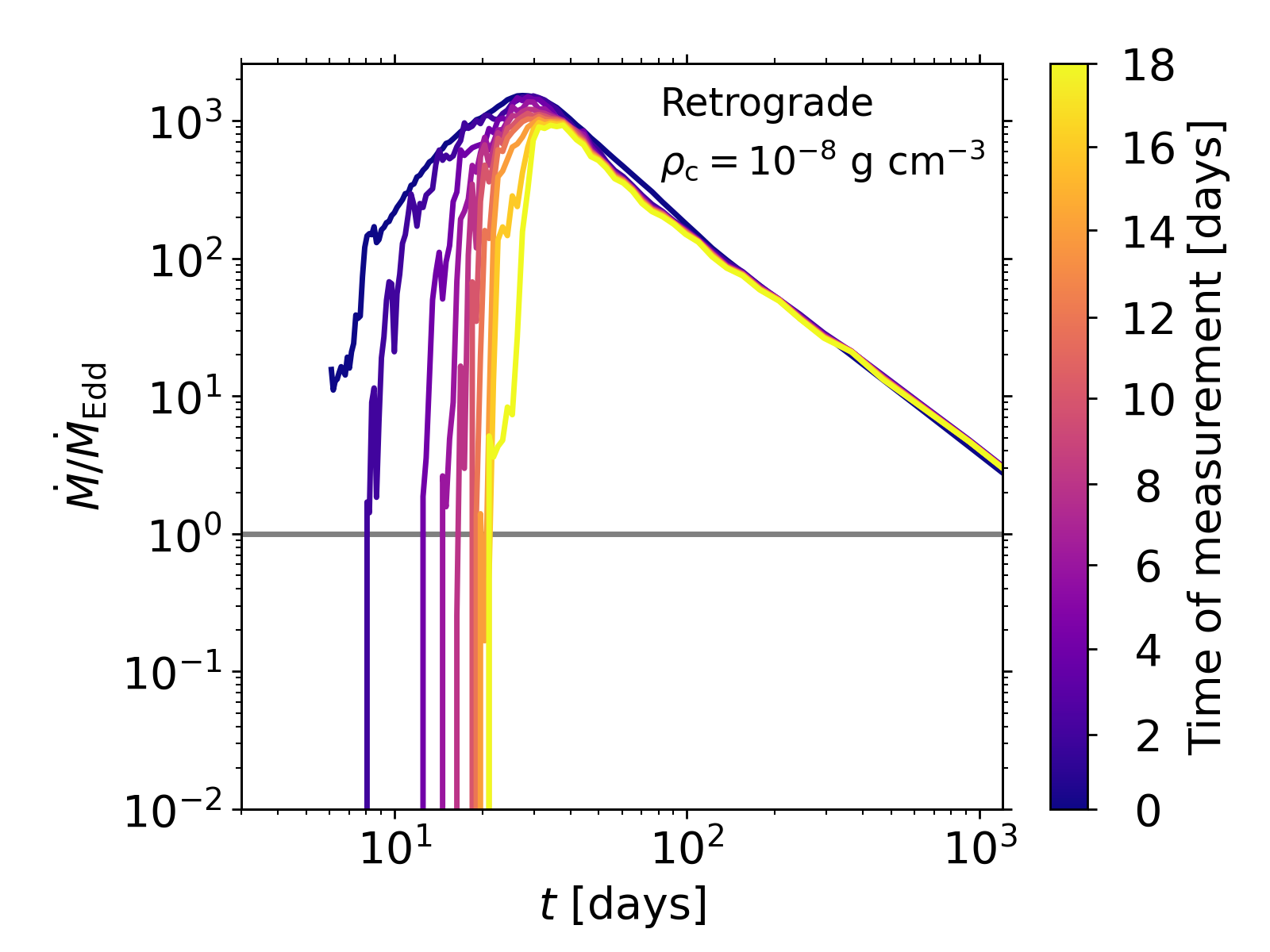}
\caption{Same as Figure~\ref{fig:stellarmass3}, but for $\mstar=10\Msol$. }
	\label{fig:stellarmass10}
\end{figure*}

\end{document}